\documentclass[twocolumn,showpacs,preprintnumbers,amsmath,%
amssymb,nofootinbib,floatfix]{revtex4}
\usepackage[english]{babel}

\usepackage{graphicx}
\usepackage{calrsfs}
\usepackage[hypertex]{hyperref}

\begin{document}
\def\be{\begin{equation}}
\def\ee{\end{equation}}
\def\ba{\begin{eqnarray}}
\def\ea{\end{eqnarray}}
\def\e{\epsilon}
\def\d{\delta}
\def\t{\tilde}
\def\o{\omega}
\def\ds{\displaystyle}
\def\ms{\mathstrut}
\def\c{\centerline}

\title[Energy spectra of gamma-rays, electrons and neutrinos]%
{Energy spectra of gamma-rays, electrons and neutrinos produced at
interactions of relativistic protons with low energy radiation}

\author{S.R.~Kelner}
\altaffiliation{Moscow Institute of Engineering Physics, Kashirskoe sh.
31, Moscow, 115409 Russia; Max-Planck-Institut f\"ur Kernphysik,
Saupfercheckweg 1, D-6917 Heidelberg, Germany } \email{skelner@rambler.ru}

\author{F.A.~Aharonian}
\altaffiliation{Dublin Institute for Advanced Studies, 31 Fitzwilliam Place, Dublin 2, Ireland;
Max-Planck-Institut f\"ur Kernphysik,
Saupfercheckweg 1, D-6917 Heidelberg, Germany}
\email{Felix.Aharonian@mpi-hd.mpg.de}

\begin{abstract}
We derived simple analytical parametrizations for energy
distributions of photons, electrons, and neutrinos
produced in interactions of relativistic protons with
an isotropic monochromatic radiation field. The results
on photomeson processes are obtained using numerical simulations
of proton-photon interactions based on the public available Monte-Carlo
code SOPHIA. For calculations of
energy spectra of electrons and positrons from the pair production
(Bethe-Heitler) process we suggest a simple formalism
based on the well-known differential cross-section of the process
in the rest frame of the proton.
The analytical presentations of energy distributions of photons and leptons
provide a simple but accurate approach for calculations
of broad-band energy spectra of gamma-rays and neutrinos in
cosmic proton accelerators located in radiation dominated environments.
\end{abstract}

\pacs{12.20.Ds, 13.20.Cz, 13.60.-r, 13.85.Qk}

\maketitle

\section{Introduction}

In astrophysical environments the density of low-energy radiation
often exceeds the density of the gas component. Under such conditions, the
interactions of ultrarelativistic protons and nuclei with radiation can
dominate over interactions with the ambient gas.
These interactions proceed through three channels:
(i) inverse Compton scattering, $p \gamma \to p \gamma^\prime$,
(ii) electron-positron pair production, $p + \gamma \to p e^+ e^-$, and
(iii) photomeson production, $p \gamma \to N + k \pi$. While the inverse Compton scattering
does not have a kinematic threshold,
the electron-positron pair production and
the photomeson production processes take place when the
energy of the photon in the rest frame of the
projectile proton exceeds $2 m_ec^2 \simeq 1\,\rm MeV$
and $m_\pi c^2 (1+m_\pi/2m_p) \simeq 145 \rm MeV$, respectively.

The process of inverse Compton scattering of protons is identical to the
inverse Compton scattering of electrons, but the energy loss-rate of protons
is suppressed by a factor of $(m_e/m_p)^4 \approx 10^{-13}$. At energies above
the threshold of production of electron-positron pairs this process is four
orders of magnitude slower compare to the losses caused by pair-production.
Therefore, generally the inverse Compton scattering does not play significant
role even in extremely dense radiation fields.

The cross-section of $(e^+,e^-)$ pair production (often
called as Bethe-Heitler cross-section) is quite large, but only a small
($\leq 2m_e/m_p$) fraction of the proton energy is converted to the
secondary electrons. The cross-section of photomeson production is
smaller, but instead a substantial (10 percent or more) fraction of the proton energy
is transferred to the secondary product. As a result, when the
proton energy exceeds the $\pi$-meson production threshold,
the hadronic interactions of protons dominate over the pair-production.

The cross-section of pair-production is calculated
with a very high accuracy using the standard routines
of quantum electrodynamics. The cross-sections of photomeson
processes are provided from accelerator experiments and
phenomenological studies. Generally,
for astrophysical applications the data obtained in fixed target
experiments with gamma-ray beams of energies from
150~MeV to 10~GeV are sufficient, especially in the case
of broad-band spectra of target photons, when the hadron-photon
interactions are contributed mainly from the region not far from
the energy threshold of the process.

The energy losses of protons in the photon fields, in particular
in the context of interactions of highest energy cosmic rays
with 2.7 K cosmic microwave background radiation (CMBR),
have been comprehensively studied by many authors
(see e.g. Refs.~\cite{Stecker,Blumenthal,Berezinsky,Chod,Stanev}). Less
attention has been given to calculations of the products energy distributions.
Partly this can be explained
by the fact that the cross-sections of secondary electrons and
gamma-rays with the ambient photons significantly exceeds
the cross-sections of interactions of protons with the same target photons.
Therefore the electrons and gamma-rays cannot leave the active regions
of their production, but rather trigger electromagnetic cascades in the
surrounding radiation and magnetic fields. The spectra of
gamma-rays formed during the cascade development are not sensitive
to the initial energy distributions, and therefore simple approximate
approaches (see e.g. Refs.~\cite{AtDer}) can provide adequate accuracies for calculations of
the characteristics of optically thick sources.
This does not concern, however, neutrinos which freely escape the
source and thus have a undistorted imprint of parent protons.
Moreover, at some specific conditions, the secondary electrons from the
pair- and photomeson production processes can cool mainly
through synchrotron radiation for which the source can be optically
thin (see e.g. \cite{FA,Susumu,Stefano,Armengaud}).
Thus the exact calculations of spectra of secondary electrons are
quite important, since the synchrotron radiation of these electrons carry
direct information about the energy spectra of accelerated protons.

The interactions of hadrons with radiation fields can be effectively
modeled by Monte Carlo simulations of characteristics of
secondary products. In particular, the SOPHIA code \cite{SOPHIA}
provides an adequate tool for the comprehensive study of
high energy properties of hadronic sources in which
interactions of ultrarelativistic protons with ambient
low-energy photons dominate over other processes. At the same time,
it is useful to have a complementary tool
for study of radiation characteristics of hadronic sources,
especially when one deals with simple scenarios, e.g.
interactions of protons with
a homogeneous and isotropic source of radiation in which the
hadronic cascades (i.e. the next generation particles)
do not play an important role. Motivated by this objective,
in this paper we develop a simple approach based on the
description of the energy distributions of final
products from the photomeson and electron-positron
pair production processes in
analytical forms which can be easily integrated in any complex
model of hadronic interactions in high energy astrophysical sources.
To a certain extent, this paper can be
considered as a continuation of our first paper \cite{Kelner1} where
we obtained analytical presentations for proton-proton interactions.

\section{Photomeson processes}

The formation of high energy gamma-rays, electrons and
neutrinos in photomeson interactions proceeds through
pro\-duc\-tion and decay of non-stable secondary products,
mainly $\pi^0$ and $\pi^\pm$-mesons:
 \be\label{ggpi0}
\gamma+p\to n_0^{}\pi^0_{}+n_+^{}\pi^+_{}+n_-^{}\pi^-_{}+\cdots
 \ee
where $n_0^{}$, $n_+^{}$ and $n_-^{}$ are the numbers of produced pions.
Hereafter we will assume that the density of the ambient medium is sufficiently low,
thus the pions decay before they interact with the surrounding gas, radiation and
magnetic fields. We will also assume that
(i) both the relativistic protons and the target low-energy photons
are isotropically distributed, and
(ii) the energy of colliding particles
 \be\label{cond}
 \e\ll m_\pi c^2\,,\quad E_p\gg m_p c^2\,,
 \ee
 where
 $\e$ is the energy of the
target photon, $m_\pi$ is the mass of $\pi$-meson
(we will assume that $m_\pi^0 = m_\pi^+$), and
$E_p$ and $m_p$ are the energy and the mass of proton.
Those conditions, which are always satisfied in astronomical
environments, allows us to obtain simple analytical presentations
for energy distributions of the final products of $\pi$- decays - photons and leptons.

\begin{figure}
\begin{center}
\includegraphics[width=0.3\textwidth,angle=-90]{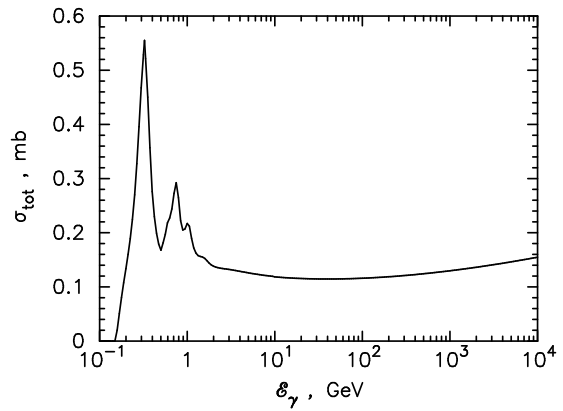}
\caption{\label{gamma_p}
\small The total cross-section of the inelastic $\gamma$-$p$ interactions as a function
of energy of gamma-ray in the proton rest frame. The calculations have been performed
using the routine of the code SOPHIA \cite{SOPHIA}.}
\end{center}
\end{figure}

The total cross-section of inelastic $\gamma$-$p$ interactions is a function of the scalar
$\mathcal{E}_\gamma\equiv(k\cdot p)/m_p$, where $k$ and $p$ are four-momenta
of the photon and proton. The scalar $\mathcal{E}_\gamma$ is simply the energy of the
photon in the proton rest frame. The total cross-section $\sigma(\mathcal{E}_\gamma)$,
calculated using the routines of the code SOPHIA \cite{SOPHIA},
is presented in Fig.~\ref{gamma_p}. Although the
production of pions dominate in the $\gamma$-$p$ interactions,
some other channels, in particular the ones leading to production of
$K$- and $\eta$-mesons, contribute noticeably (up to 10 to 20 \%)
to the overall production of photons and leptons.
These channels are taken into account
in our calculations presented below. They are based on the
SOPHIA code which allows simulations of all
important processes belongs to $\gamma$-$p$ interactions.

\subsection{Production of gamma-rays}

The inclusive cross-section of production of $\pi^0_{}$-mesons
\be\label{eq1}
d\sigma_{\pi^0}=S({\bf k},{\bf p},{\bf p}_{\pi^0})\,d^3p_{\pi^0}
\ee
depends on the momenta ${\bf k},\,{\bf p},\,{\bf p}_{\pi^0}$ of the photon, proton and
$\pi^0_{}$-meson, respectively. Let's denote by
\be\label{eq2}
dw=W({\bf p}_{\pi^0},{\bf p}_{\gamma})\,d^3p_{\gamma}
\ee
the probability of decay of a $\pi^0_{}$-meson with momentum
${\bf p}_{\pi^0}$ into a gamma-ray photon of momentum ${\bf p}_{\gamma}$
in the volume $d^3p_{\gamma}$ of the momentum space. Then
\be\label{eq3}
d\sigma_{\gamma}=2\,d^3p_{\gamma}\int S({\bf k},{\bf p},{\bf p}_{\pi^0})\,
W({\bf p}_{\pi^0},{\bf p}_{\gamma})\,d^3p_{\pi^0}
\ee
can be treated as the inclusive cross-section
of $\gamma$-rays production through the chain
$\gamma+p\to n_0^{}\pi^0_{}+\cdots\to 2\,n_0^{}\gamma+\cdots$.

For isotropic distributions of initial particles, the final products will
be isotropically distributed as well. Therefore for the determination of the energy spectra
of $\gamma$-rays we can use the inclusive cross-section given by Eq.(\ref{eq3}) but
integrated over the gamma-rays emission angles:
\be\label{eq5}
d\sigma_\gamma\equiv G_\gamma(E_p,\e,\cos\theta,E_\gamma)\,\frac{dE_\gamma}{E_p}
\ee
The function $G_\gamma$ depends on $E_p$, $\e$, $E_\gamma$,
and the angle $\theta$ between the momenta of colliding proton and photon.
The corresponding differential interaction rate is
\be\label{eq6}
dw_\gamma=c\,(1-\cos\theta)\, G_\gamma(E_p,\e,\cos\theta,E_\gamma)\,\frac{dE_\gamma}{E_p}\,.
\ee

Let $f_p(E_p)$ and $f_{\rm ph}(\e)$ be functions characterizing the energy distributions
of initial protons and photons, i.e. $f_p(E_p)\,dE_p$ and $f_{\rm ph}(\e)\,d\e$
are the numbers of protons and photons per $1\,{\rm cm}^3_{}$ in the energy intervals
$dE_p$ and $d \e$. respectively.
Since it is assumed that the target photons are isotropically 
distributed, their angular distribution is described as 
$d\Omega/4\pi$. Then the production rate of $\gamma$-rays
(i.e. number of $\gamma$-rays in the energy interval
$(E_\gamma, E_\gamma+d E_\gamma)$ per sec, per $1\,{\rm cm}^3_{}$) 
can be obtained after integration of Eq.~(\ref{eq6}) over 
energies of protons and target photons, as well as over the  solid angle $d\Omega$:

\ba\label{eq7}
&d N_\gamma(E_\gamma)=
\ds dE_\gamma\!\int\!\,\frac{dE_p}{E_p}\,\frac{d\Omega}{4\pi}\,d\e\,f_p(E_p)\,f_{\rm ph}(\e)
\times\nonumber&\\[4pt]
&c\,(1-\cos\theta)\, G_\gamma(E_p,\e,\cos\theta,E_\gamma)\,.&
\ea

Let's introduce the function $\Phi_\gamma$ defined as
\be\label{eq8}
\Phi_\gamma(\eta,x)\equiv
\int\! c\,(1-\cos\theta)\, G_\gamma(E_p,\e,\cos\theta,E_\gamma)
\,\frac{d\Omega}{4\pi}\,,
\ee
where
\be\label{eq9}
\eta=\frac{4\,\e E_p}{m_p^2c^4}\,,\qquad x=\frac{E_\gamma}{E_p}.
\ee
Then Eq.(\ref{eq7}) can be written in the following form
\be\label{eq10}
\frac{dN_\gamma}{dE_\gamma}=
\int\! f_p(E_p)\,f_{\rm ph}(\e)\; \Phi_\gamma\!\left(\eta,\,x\right) \frac{dE_p}{E_p}\,d\e\,.
\ee
Note that $\Phi_\gamma$ can be treated as a function of two (but not three)
variables. This is connected with the fact that,
as we assume, at $E_p\to\infty$ and
$\e\to0$, $E_\gamma\to\infty$, and for fixed $x$ and $\eta$,
the function $\Phi_\gamma$ should have a certain limit. In other words,
at the rescaling
\be\label{eq11}
E_p\to\lambda E_p\,\quad E_\gamma\to\lambda E_\gamma\,,\quad
\e\to\e/\lambda\,,
\ee
where $\lambda$ is an arbitrary number ($\lambda>1$),
the function $\Phi_\gamma$
is not changed, therefore the following relation takes place.
\ba\label{eq12}
&\int (1-\cos\theta) \,G_\gamma(\lambda E_p,\e/\lambda,\cos\theta,\lambda E_\gamma)\,d\Omega
=\nonumber&\\[4pt]
&=\int (1-\cos\theta)\,G_\gamma(E_p,\e,\cos\theta,E_\gamma)\,d\Omega\,.
\ea
%
Numerical calculations based on the code SOPHIA~\cite{SOPHIA} show that already at
\be
\e<10^{-3}\, m_\pi c^2\,,\qquad E_p>10^3\, m_p c^2
\ee
the relation given by Eq.(\ref{eq12}) is readily fulfilled.

Some features of the energy spectra of $\gamma$-rays can be understood from the
analysis of the kinematics of the process. In particular,  
for production of a single $\pi^0_{}$-meson
the following condition should be satisfied
\be\label{kin1}
2\,\e E_p\,(1-\beta_p\cos\theta)>(2\,m_\pi m_p+m_\pi^2)\,c^4\,,
\ee
where
$\beta_p$ -- is the proton speed (in the units of $c$).

Since the protons are ultrarelativistic, we will assume $\beta_p=1$.
If the condition of Eq.(\ref{kin1}) is not satisfied, then the interaction rate given
by Eq.(\ref{eq6}) is equal to zero. Therefore, for the case of
$4\e E_p\le (2m_\pi m_p+m_\pi^2)c^4$,
the function $\Phi_\gamma=0$. The integration of Eq.(\ref{eq10}) should
be performed over the region\footnote{For determination of the region
allowed by kinematics, we will assume $m_{\pi^\pm}=m_{\pi^0}=0.137\,$GeV.
Since on the boarder of this region the function $\Phi_\gamma=0$, this approximation
does not affect the accuracy of numerical calculations.}
\be\label{kin2}
\eta\ge\eta_0^{}\equiv 2\,\frac{m_\pi}{m_p}+\frac{m_\pi^2}{m_p^2}\approx 0.313\,,
\ee
As it follows from kinematics of production
of a single pion, the energy of the latter appears within
%
\be\label{kin3}
E_{\pi\min}\le E_{\pi}\le E_{\pi\max}\,,
\ee
where
\be\label{kin4}
E_{\pi\max}=E_p\,x_+^{}\,,\qquad E_{\pi\min}=E_p\,x_-^{}
\ee
are the maximum and minimum energies respectively, with
\be\label{kin5}
x_{\pm}^{}=\frac{1}{2(1+\eta)}\,\left[\eta+r^2\pm\sqrt{(\eta-r^2-2r)\,(\eta-r^2+2r)}\right],
\ee
where $r=m_\pi/m_p\approx 0.146$.

Let's consider now the general case when the total mass of particles
produced in proton-photon interactions is $M+m_\pi$. For example, the single-pion
production implies $M=m_p$, while in the case of two-pion production $M=m_p+m_\pi$,
{\it etc}. In this case the maximum and minimum energies of pions are
\be\label{kin6}
\tilde E_{\pi\max}=E_p\,\tilde x_+^{}\,,\qquad \tilde E_{\pi\min}=E_p\,\tilde x_-^{}\,.
\ee
with
\ba\label{kin7}
&\ds\tilde x_{\pm}^{}=\frac{1}{2(1+\eta)}\,\left[\eta+r^2+1-R^2\pm\phantom{\frac aa}\right. \nonumber&\\[4pt]
&\left.\pm\sqrt{(\eta+1-(R+r)^2)\,(\eta+1-(R-r)^2)}\right],&
\ea
where $R=M/m_p$. In particular, for $R=1$ Eqs.~(\ref{kin5}) and (\ref{kin7})
coincide. For $R>1$, we have the following inequalities
\be\label{kin7a}
E_{\pi\min}<\tilde E_{\pi\min}\,,\qquad E_{\pi\max}>\tilde E_{\pi\max}\,.
\ee

The decay of ultrarelativistic $\pi^0_{}$-mesons with energy distribution $J_\pi(E_\pi)$
within the limits given by Eq.(\ref{kin3}), results in the energy spectrum of $\gamma$-rays
\be\label{kin8}
\frac{dN_\gamma}{dE_\gamma}=2\int\limits_{E_1}^{E_{\pi\max}}\frac{dE_\pi}{E_\pi}\,J_\pi(E_\pi)\,,
\ee
where
\be\label{kin9}
E_1=\max\!\left(E_\gamma,\frac{m_\pi^2c^4}{4E_\gamma}, E_{\pi\min}\right), \quad E_\gamma<E_{\pi\max}\,.
\ee
Below we will be interested in $\gamma$-rays with energy $E_\gamma>m_\pi/2$.
While in the energy range
$E_\gamma<E_{\pi\min}$ the dif\-fe\-ren\-tial spectrum of $\gamma$-rays is flat,
\be\label{kin10}
\frac{dN_\gamma}{dE_\gamma}=2\int\limits_{E_{\pi\min}}^{E_{\pi\max}}\frac{dE_\pi}{E_\pi}\,J_\pi(E_\pi)\,,
\ee
within the interval $E_{\pi\min}<E_\gamma<E_{\pi\max}$ the spectrum decreases with energy,
\be\label{kin11}
\frac{dN_\gamma}{dE_\gamma}=2\int\limits_{E_\gamma}^{E_{\pi\max}}\frac{dE_\pi}{E_\pi}\,J_\pi(E_\pi)\,.
\ee
The above quantitative features of the spectrum of $\gamma$-rays
appear quite useful for the choice of approximate analytical presentations.
The results of numerical calculations of the function $\Phi_\gamma$
based on simulations using the code SOPHIA \cite{SOPHIA} can be approximated,
with an accuracy better than 10\% by simple analytical formulae. Namely, in the range
 $x_-<x<x_+$
\[
\Phi_\gamma(\eta,x)=B_\gamma\,\exp\!\left\{-s_\gamma
\left[\ln\!\left(\frac{x}{x_-^{}}\right)\right]^{\d_\gamma}\right\}\times\\[-8pt]
\]
\be\label{gamma1}
\times\left[\ln\!\left(\frac{2}{1+y^2}\right)
\right]^{2.5+0.4\ln(\eta/\eta_0^{})}\,,
\ee
where
\be\label{gamma2}
y=\frac{x-x_-^{}}{x_+^{}-x_-^{}}\,.
\ee
At low energies, $x<x_-$, the spectrum does not depend on $x$,
\be\label{gamma3}
\Phi_\gamma(\eta,x)=B_\gamma\,\big(\ln2\big)^{2.5+0.4\ln(\eta/\eta_0^{})}\,.
\ee
Finally in the range $x>x_+$ the function $\Phi_\gamma=0$.

\begin{figure*}
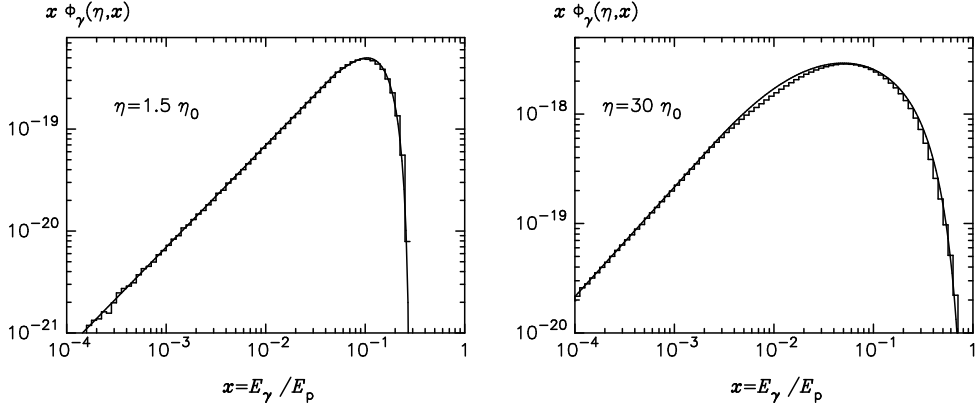

\begin{center}
\mbox{\includegraphics[width=0.3\textwidth,angle=-90]{fig2a.eps}\qquad
\includegraphics[width=0.3\textwidth,angle=-90]{fig2b.eps}}
\caption{\label{soph_gam}\small Gamma-ray spectra produced in photomeson interactions
calculated for two values of $\eta=4 \e E_p/m_p^2c^4$. Solid lines are calculated using
the analytical presentations given by Eq.(\ref{eq10}), the histograms are from
simulations using the SOPHIA code.}
\end{center}
\end{figure*}

All three parameters $B_\gamma$, $s_\gamma$ and $\d_\gamma$ used in this presentation
are functions of $\eta$. The numerical values of these parameters are shown in
Table~\ref{Table1}.
At $\eta/\eta_0=1$, i.e. at the threshold of $\pi^0_{}$-meson production, $B_\gamma=0$.
In Fig.~\ref{soph_gam} we show the functions
$x\,\Phi_\gamma(\eta,x)$ obtained with the code SOPHIA (histograms) and
the analytical presentations given by Eqs.~(\ref{gamma1}) and (\ref{gamma3})
for two values of $\eta$.

Eq.(\ref{eq10}) provides a simple approach for calculations of $\gamma$-ray spectra for
arbitrary energy distributions of ultra\-re\-la\-ti\-vi\-stic protons and ambient photons. The
parameters $B_\gamma$, $s_\gamma$ and $\d_\gamma$ are
quite smooth functions of $\eta$, thus for calculations of these parameters at
intermediate values of $\eta$ one can use linear interpolations
of the numerical results presented in Table\-~\ref{Table1}. Note that
for interactions of protons with 2.7~K CMBR, the results presented
in Table\-~\ref{Table1} allow calculations of $\gamma$-ray spectra up to
$\sim 10^{21}\,\textrm{eV}$.

\begin{table}
\caption{\label{Table1} Numerical values of parameters $B_\gamma$, $s_\gamma$
and $\d_\gamma$ characterizing the $\gamma$-ray spectra given by Eq.(\ref{eq10}).}
\begin{tabular}{cccc}
\hline
$\eta/\eta_0^{}$ &$s_\gamma$ &$\d_\gamma$& $B_\gamma\,,\,
{\rm cm}^3/{\rm s}\phantom{\ds \frac ss}\!$\\
\hline
1.1&0.0768 &0.544&$2.86\cdot10^{-19}\phantom{\ds \frac ss}\!$\\
\hline
1.2&0.106&0.540&$2.24\cdot10^{-18}\phantom{\ds \frac ss}\!$\\
\hline
1.3&0.182&0.750&$5.61\cdot10^{-18}\phantom{\ds \frac ss}\!$\\
\hline
1.4&0.201&0.791&$1.02\cdot10^{-17}\phantom{\ds \frac ss}\!$\\
\hline
1.5&0.219&0.788&$1.60\cdot10^{-17}\phantom{\ds \frac ss}\!$\\
\hline
1.6&0.216&0.831&$2.23\cdot10^{-17}\phantom{\ds \frac ss}\!$\\
\hline
1.7&0.233&0.839&$3.10\cdot10^{-17}\phantom{\ds \frac ss}\!$\\
\hline
1.8&0.233&0.825&$4.07\cdot10^{-17}\phantom{\ds \frac ss}\!$\\
\hline
1.9&0.248&0.805&$5.30\cdot10^{-17}\phantom{\ds \frac ss}\!$\\
\hline
2.0&0.244&0.779&$6.74\cdot10^{-17}\phantom{\ds \frac ss}\!$\\
\hline
3.0&0.188&1.23&$1.51\cdot10^{-16}\phantom{\ds \frac ss}\!$\\
\hline
4.0&0.131&1.82&$1.24\cdot10^{-16}\phantom{\ds \frac ss}\!$\\
\hline
5.0&0.120&2.05&$1.37\cdot10^{-16}\phantom{\ds \frac ss}\!$\\
\hline
6.0&0.107&2.19&$1.62\cdot10^{-16}\phantom{\ds \frac ss}\!$\\
\hline
7.0&0.102&2.23&$1.71\cdot10^{-16}\phantom{\ds \frac ss}\!$\\
\hline
8.0&0.0932&2.29&$1.78\cdot10^{-16}\phantom{\ds \frac ss}\!$\\
\hline
9.0&0.0838&2.37&$1.84\cdot10^{-16}\phantom{\ds \frac ss}\!$\\
\hline
10.0&0.0761&2.43&$1.93\cdot10^{-16}\phantom{\ds \frac ss}\!$\\
\hline
20.0&0.107&2.27&$4.74\cdot10^{-16}\phantom{\ds \frac ss}\!$\\
\hline
30.0&0.0928&2.33&$7.70\cdot10^{-16}\phantom{\ds \frac ss}\!$\\
\hline
40.0&0.0772&2.42&$1.06\cdot10^{-15}\phantom{\ds \frac ss}\!$\\
\hline
100.0&0.0479&2.59&$2.73\cdot10^{-15}\phantom{\ds \frac ss}\!$\\
\hline
\end{tabular}
\end{table}

\subsection{Production of electrons and neutrinos}

The production of leptons in proton-photon inte\-rac\-tions is
dominated by the decay of secondary charged pions. In analogy with Eq.(\ref{eq10}),
the spectrum of each type of leptons can be presented in the form
\ba\label{lepton1}
& N_l(E_l)\,dE_l=\nonumber&\\[4pt]
&=\ds dE_l\int\! f_p(E_p)\,f_{\rm ph}(\e)\; \Phi_l\!\left(\eta,\,x\right)
\frac{dE_p}{E_p}\,d\e\,,&
\ea
where $\eta$ is determined in Eq.(\ref{eq9}), $x=E_l^{}/E_p^{}$
and $l$ implies one of the following symbols: $e^+_{},e^-_{},\nu_\mu^{},
\bar\nu_\mu^{},\nu_e^{},\bar \nu_e^{}$.

As in the case of $\gamma$-rays, the energy range of leptons, for the fixed
values of $E_p$ and $\e$ is determined by kinematics. We obtained analytical
forms for the energy spectra for all lepton types, $\Phi_l$,
using numerical results of simulations the SOPHIA code.
In the range of $x'_-<x<x'_+$
\[
\Phi_l(\eta,x)=B_l^{}\,\exp\!\left\{-s_l^{}
\left[\ln\!\left(\frac{x}{x'_-}\right)\right]^{\d_l^{}}\right\}\times\\[-8pt]
\]
\be\label{pggq1}
\left[\ln\!\left(\frac{2}{1+y'^2}\right)
\right]^{\psi}\,,
\ee
where
\be\label{pggq1a}
y'=\frac{x-x'_-}{x'_+-x'_-}\,.
\ee
In range of $x<x'_-$, the function $\Phi_l$ does not depend on $x$:
\be\label{pggq1b}
\Phi_l(\eta,x)=B_l^{}\left(\ln 2\right)^{\psi}\,.
\ee
For $x\ge x'_+$ the function $\Phi_l=0.$
The analytical presentation in the form of Eq.(\ref{pggq1})
contains three parameters $s_l^{}$, $\d_l^{}$ and $B_l^{}$
which themselves are functions of $\eta$. The numerical values of these parameters
obtained with the method of least squares are tabulated in
Table~\ref{Table2} (for $e^+_{}$, $\bar \nu_\mu^{}$, $\nu_\mu^{}$ and $\nu_e^{}$)
and Table~\ref{Table3} (for $e^-_{}$ and $\bar \nu_e^{}$).
The values of $x'_\pm$ and $\psi$ are given below for
each type of leptons.

\begin{table*}
\caption{\label{Table2} Numerical values of parameters $s_l^{}$, $\delta_l^{}$, and $B_l^{}$
for $e^+_{}$, ${\bar\nu}_{\mu}^{}$, $\nu_{\mu}^{}$, $\nu_e^{}$.}
\begin{tabular}{c|ccc|ccc|ccc|ccc}
\hline
$\eta/\eta_0^{}\;$ &$s_{e^+}$ &$\d_{e^+}$& $B_{e^+}\,,\,{\rm cm}^3/{\rm s}\phantom{\ds \frac ss}\!$
&$s_{{\bar\nu}_{\mu}}$ &$\d_{{\bar\nu}_{\mu}}$&
$B_{{\bar\nu}_{\mu}}\,,\,{\rm cm}^3/{\rm s}\phantom{\ds \frac ss}\!$
&$s_{{\nu}_{\mu}}$ &$\d_{{\nu}_{\mu}}$&
$B_{{\nu}_{\mu}}\,,\,{\rm cm}^3/{\rm s}\phantom{\ds \frac ss}\!$
 &$s_{\nu_e}$ &$\d_{\nu_e}$& $B_{\nu_e}\,,\,{\rm cm}^3/{\rm s}\phantom{\ds \frac ss}\!$\\
\hline
1.1 &0.367&3.12&$8.09\cdot 10^{-19}$&0.365&3.09&$8.09\cdot 10^{-19}$&0.0&0.0&$1.08\cdot 10^{-18}$
&0.768&2.49&$9.43\cdot 10^{-19}\phantom{\ds \frac ss}\!$\\
1.2 &0.282&2.96&$7.70\cdot 10^{-18}$&0.287&2.96&$7.70\cdot 10^{-18}$&0.0778&0.306&$9.91\cdot 10^{-18}$
&0.569&2.35&$9.22\cdot 10^{-18}\phantom{\ds \frac ss}\!$\\
1.3&0.260&2.83&$2.05\cdot 10^{-17}$ &0.250&2.89&$1.99\cdot 10^{-17}$&0.242&0.792&$2.47\cdot 10^{-17}$
&0.491&2.41&$2.35\cdot 10^{-17}\phantom{\ds \frac ss}\!$\\
1.4&0.239&2.76&$3.66\cdot 10^{-17}$&0.238&2.76&$3.62\cdot 10^{-17}$&0.377&1.09&$4.43\cdot 10^{-17}$
&0.395&2.45&$4.20\cdot 10^{-17}\phantom{\ds \frac ss}\!$\\
1.5&0.224&2.69&$5.48\cdot 10^{-17}$&0.220&2.71&$5.39\cdot 10^{-17}$&0.440&1.06&$6.70\cdot 10^{-17}$
&0.31&2.45&$6.26\cdot 10^{-17}\phantom{\ds \frac ss}\!$\\
1.6&0.207&2.66&$7.39\cdot 10^{-17}$&0.206&2.67&$7.39\cdot 10^{-17}$ &0.450&0.953&$9.04\cdot 10^{-17}$
&0.323&2.43&$8.57\cdot 10^{-17}\phantom{\ds \frac ss}\!$\\
1.7&0.198&2.62&$9.52\cdot 10^{-17}$&0.197&2.62&$9.48\cdot 10^{-17}$&0.461&0.956&$1.18\cdot 10^{-16}$
&0.305&2.40&$1.13\cdot 10^{-16}\phantom{\ds \frac ss}\!$\\
1.8&0.193&2.56&$1.20\cdot 10^{-16}$ &0.193&2.56&$1.20\cdot 10^{-16}$ &0.451&0.922&$1.32\cdot 10^{-16}$
&0.285&2.39&$1.39\cdot 10^{-16}\phantom{\ds \frac ss}\!$\\
1.9&0.187&2.52&$1.47\cdot 10^{-16}$&0.187&2.52&$1.47\cdot 10^{-16}$&0.464&0.912&$1.77\cdot 10^{-16}$
&0.270&2.37&$1.70\cdot 10^{-16}\phantom{\ds \frac ss}\!$\\
2.0&0.181&2.49&$1.75\cdot 10^{-16}$&0.178&2.51&$1.74\cdot 10^{-16}$&0.446&0.940&$2.11\cdot 10^{-16}$
&0.259&2.35&$2.05\cdot 10^{-16}\phantom{\ds \frac ss}\!$\\
3.0&0.122&2.48&$3.31\cdot 10^{-16}$&0.123&2.48&$3.38\cdot 10^{-16}$&0.366&1.49&$3.83\cdot 10^{-16}$
&0.158&2.42&$3.81\cdot 10^{-16}\phantom{\ds \frac ss}\!$\\
4.0&0.106&2.50&$4.16\cdot 10^{-16}$&0.106&2.56&$5.17\cdot 10^{-16}$&0.249&2.03&$5.09\cdot 10^{-16}$
&0.129&2.46&$4.74\cdot 10^{-16}\phantom{\ds \frac ss}\!$\\
5.0&0.0983&2.46&$5.57\cdot 10^{-16}$&0.0944&2.57&$7.61\cdot 10^{-16}$&0.204&2.18&$7.26\cdot 10^{-16}$
 &0.113&2.45&$6.30\cdot 10^{-16}\phantom{\ds \frac ss}\!$\\
6.0&0.0875&2.46&$6.78\cdot 10^{-16}$&0.0829&2.58&$9.57\cdot 10^{-16}$&0.174&2.24&$9.26\cdot 10^{-16}$
&0.0996&2.46&$7.65\cdot 10^{-16}\phantom{\ds \frac ss}\!$\\
7.0&0.0830&2.44&$7.65\cdot 10^{-16}$&0.0801&2.54&$1.11\cdot 10^{-15}$&0.156&2.28&$1.07\cdot 10^{-15}$
&0.0921&2.46&$8.61\cdot 10^{-16}\phantom{\ds \frac ss}\!$\\
8.0&0.0783&2.44&$8.52\cdot 10^{-16}$&0.0752&2.53&$1.25\cdot 10^{-15}$&0.140&2.32&$1.19\cdot 10^{-15}$
&0.0861&2.45&$9.61\cdot 10^{-16}\phantom{\ds \frac ss}\!$\\
9.0 & 0.0735&2.45&$9.17\cdot 10^{-16}$&0.0680&2.56&$1.36\cdot 10^{-15}$&0.121&2.39&$1.29\cdot 10^{-15}$
&0.0800&2.47&$1.03\cdot 10^{-15}\phantom{\ds \frac ss}\!$\\
10.0&0.0644&2.50&$9.57\cdot 10^{-16}$&0.0615&2.60&$1.46\cdot 10^{-15}$&0.107&2.46&$1.40\cdot 10^{-15}$
&0.0723&2.51&$1.10\cdot 10^{-15}\phantom{\ds \frac ss}\!$\\
30.0&0.0333&2.77&$3.07\cdot 10^{-15}$&0.0361&2.78&$5.87\cdot 10^{-15}$&0.0705&2.53&$5.65\cdot 10^{-15}$
&0.0411&2.70&$3.55\cdot 10^{-15}\phantom{\ds \frac ss}\!$\\
100.0&0.0224&2.86&$1.58\cdot 10^{-14}$&0.0228&2.88&$3.10\cdot 10^{-14}$&0.0463&2.62&$3.01\cdot 10^{-14}$
&0.0283&2.77&$1.86\cdot 10^{-14}\phantom{\ds \frac ss}\!$\\
\end{tabular}
\end{table*}

\subsubsection{Energy spectra of positrons, muon antineutrinos and electrons neutrinos}

\begin{figure*}
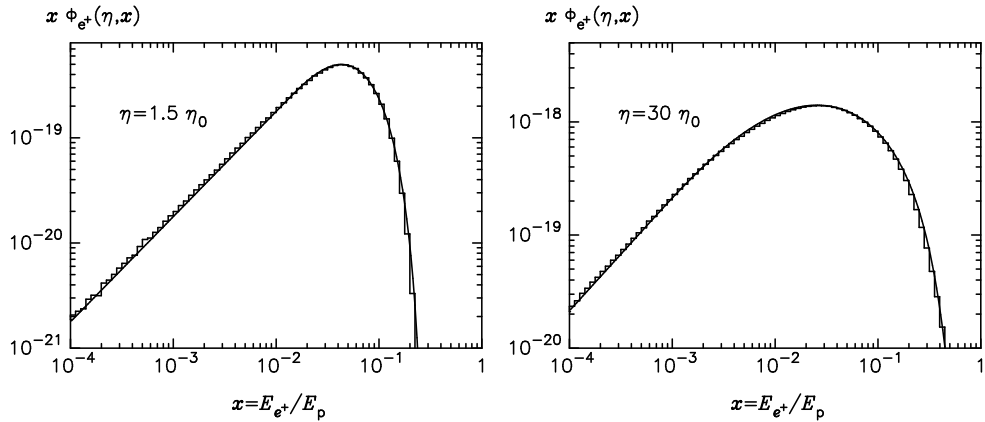

\begin{center}
\mbox{\includegraphics[width=0.31\textwidth,angle=-90]{fig3a.eps}\quad
\includegraphics[width=0.31\textwidth,angle=-90]{fig3b.eps}}
\caption{\label{soph_posi}\small Energy spectra of positrons produced in photomeson
interactions calculated for two values of $\eta=4 \e E_p/m_p^2c^4$. Solid lines are
calculated using the analytical presentations, the histograms
are from Monte-Carlo simulations using the SOPHIA code.}
\end{center}
\end{figure*}

\begin{figure*}
\begin{center}
\mbox{\includegraphics[width=0.31\textwidth,angle=-90]{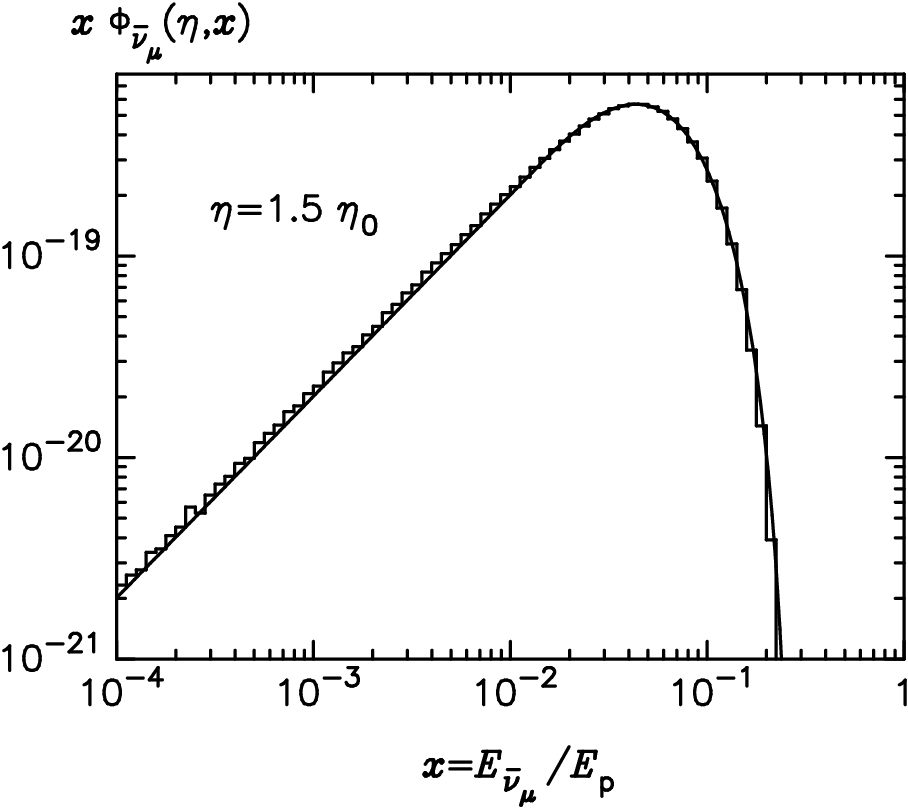}\quad
\includegraphics[width=0.31\textwidth,angle=-90]{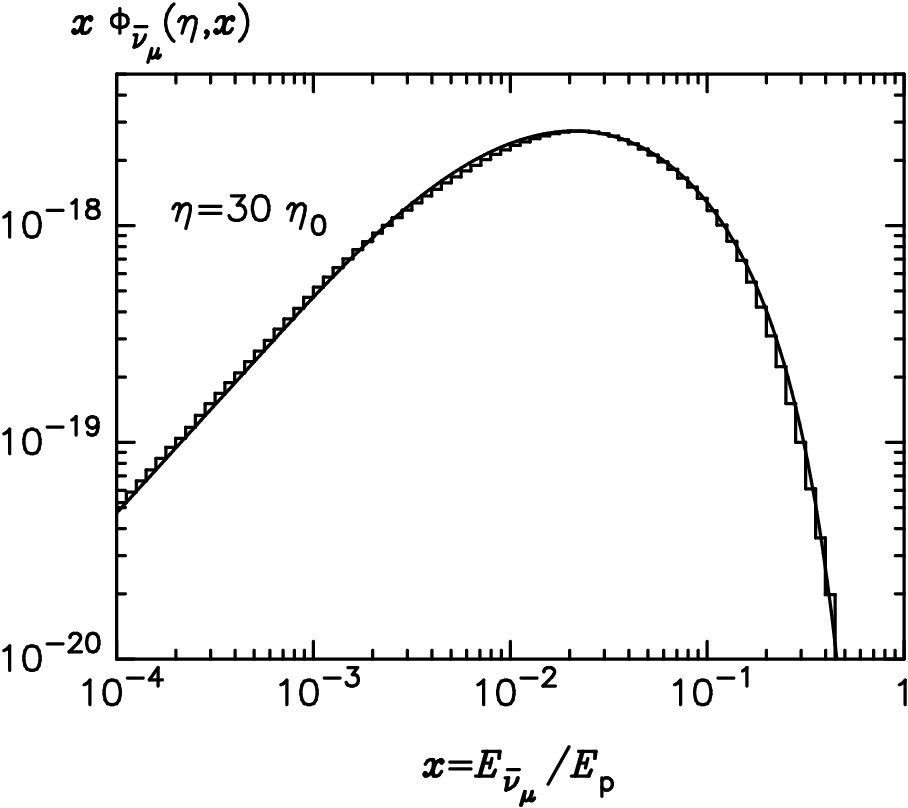}}
\caption{\label{soph_anumu}\small The same as in Fig.~\ref{soph_posi} but for muon
antineutrinos.}
\end{center}
\end{figure*}

For $e^+_{}$, $\bar \nu_\mu^{}$ and $\nu_e^{}$, the parameter $\psi$
is presented in the form
\be
\psi=2.5+1.4\ln\!\left(\frac{\eta}{\eta_0^{}}\right),
\label{psi-lepton1}
\ee
with
\be
x'_-=\frac{x_-}4 \quad \rm and \quad x'_+=x_+,
\ee
where $x_+$ and $x_-$ are determined from Eq.(\ref{kin5}); $\eta_0^{}$ is defined
in Eq.(\ref{kin2}). Note that here $\psi$ is different than the relevant function
in Eq.(\ref{gamma1}). In Fig.~\ref{soph_posi} and Fig.~\ref{soph_anumu}
the analytical presentations of distributions $x\,\Phi_{e^+}$ and
$x\,\Phi_{\bar \nu_\mu}$ are compared with Monte-Carlo simulations based on the SOPHIA code.

In the range $\eta<4\,\frac{m_\pi}{m_p}+4\left(\frac{m_\pi}{m_p}\right)^2=2.14\,\eta_0$, only
a single $\pi^+_{}$-meson can be produced. It decays to $\pi^+_{}\to \mu^+_{}\nu_\mu^{}$.
The positrons and muon antineutrinos are produced from the decay
$\mu^+_{} \to e^+_{}\bar{\nu}_\mu\nu_e$. Since the spectra of
$e^+_{}$ and $\bar{\nu}_\mu$ from the decay of $\mu^+_{}$ coincide (see e.g. \cite{Gaisser}),
the parameters in the Tables ~\ref{Table2} calculated for
$e^+_{}$ and $\bar{\nu}_\mu$ for small $\eta$ should be identical.
The slight difference
at $\eta<2 \eta_0$ is explained by fluctuations related to the
the statistical character of simulations.

At $\eta>2.14\,\eta_0$, a new channel is opened for production of $\bar{\nu}_\mu$,
because of production of $\pi^-_{}$-mesons and their decay:
$\pi^-_{}\to \mu^-_{}\bar{\nu}_\mu$. Therefore for large values of $\eta$
the parameters characterizing $e^+_{}$ and
$\bar{\nu}_\mu$ differ significantly (see Table~\ref{Table2}). This can be seen
from comparison of results presented in right panels of Fig.~\ref{soph_posi} and
\ref{soph_anumu}; for $\eta=30$, $x\,\Phi_{\bar{\nu}_\mu}$
significantly exceeds $x\,\Phi_{e^+}$.

\begin{figure*}
\begin{center}
\mbox{\includegraphics[width=0.31\textwidth,angle=-90]{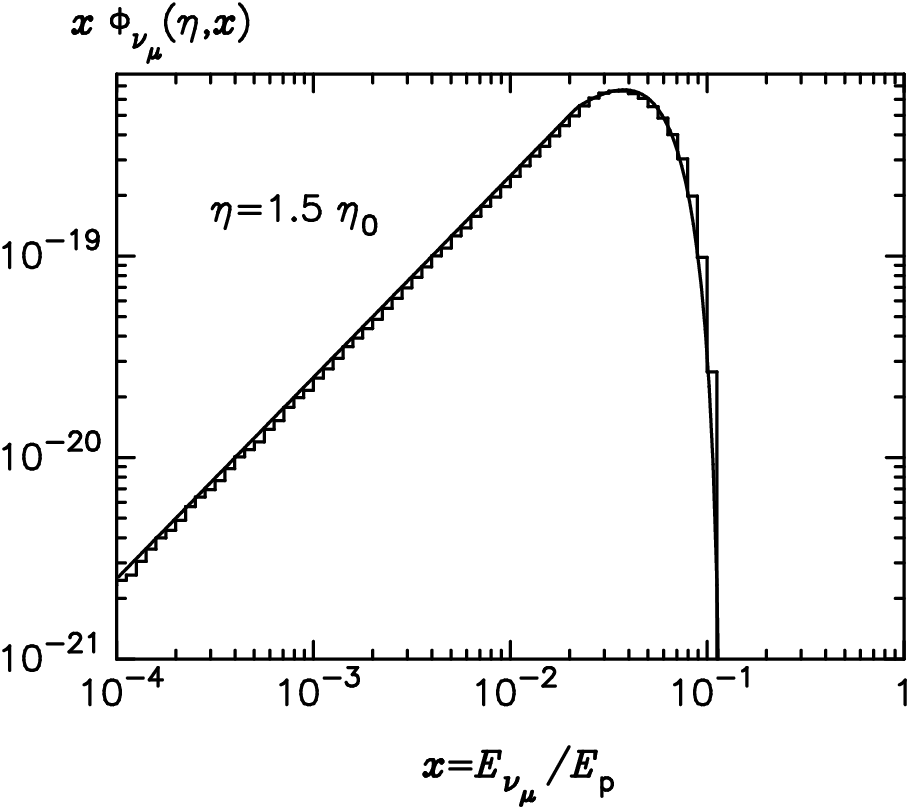}\quad
\includegraphics[width=0.31\textwidth,angle=-90]{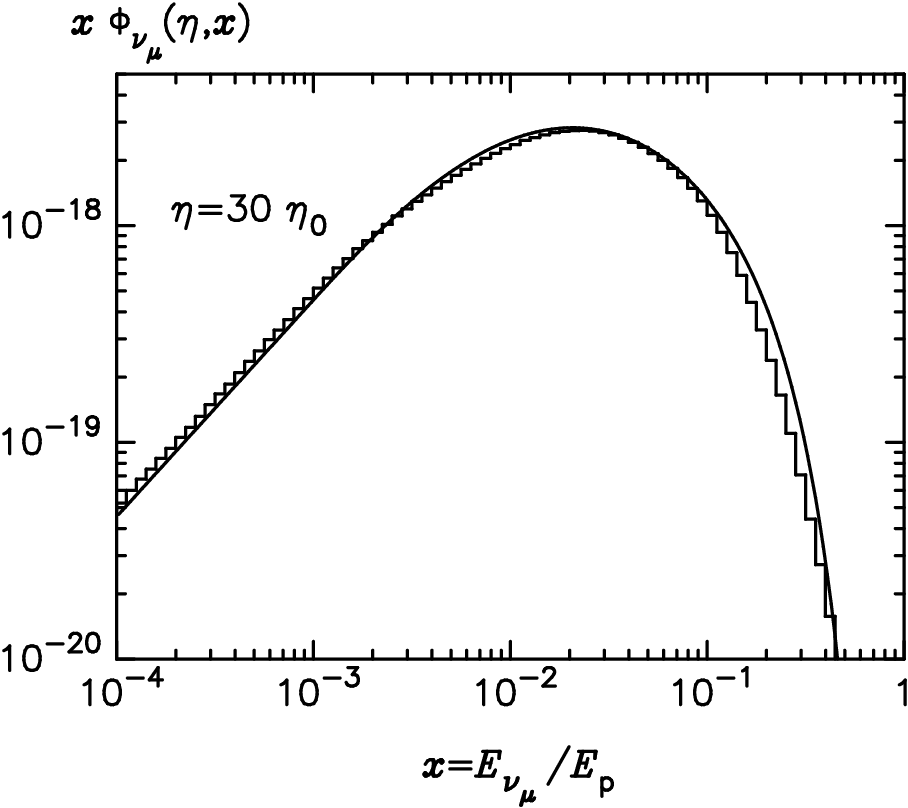}}
\caption{ \label{soph_numu}\small The same as in Fig.~\ref{soph_posi}
but for muon neutrinos.}
\end{center}
\end{figure*}

\subsubsection{Muon neutrinos}

The distribution for $\nu_\mu^{}$ is described by Eq.(\ref{pggq1}) with
the same function $\psi$ as for $e^+_{}$, $\bar \nu_\mu^{}$ and $\nu_e^{}$,
given by Eq.(\ref{psi-lepton1}), but with different parameters $x'_{\pm}$:
\be
x'_+=\left\{
\begin{array}{lc}
0.427\,x_+\,,& \rho<2.14\,,\\[3pt]
(0.427+0.0729\,(\rho-2.14))\,x_+\,,& 2.14<\rho<10\,,\\[3pt]
x_+\,,& \rho>10\,,\\[3pt]
\end{array}
\right.
\ee
where $\rho=\eta/\eta_0$, and
\be
x'_-=0.427\,x_-\,.
\ee
The difference of values of $x_{\pm}^{}$ for $\bar{\nu}_\mu$ and ${\nu}_\mu$ appears for
the following reason. At the decay $\pi^+_{}\to \mu^+_{}\bar{\nu}_\mu$
the maximum energy of $\bar{\nu}_\mu$ is equal to
$(1-m_\mu^2/m_\pi^2)\,E_\pi\approx 0.427\,E_\pi$,
where $m_\mu$ is the mass of muon. On the other hand, the maximum energy of
$\bar{\nu}_\mu$, produced at decay of the muon is equal to $E_\pi$. With an increase of
the parameter $\eta$, $\pi^-_{}$-mesons start to be produced, the decay of which
leads to ${\nu}_\mu$ with maximum energy comparable to the energy of the pion
(a detailed discussion of these questions can be found in \cite{Kelner1}).

In Fig.~\ref{soph_numu} the function $x\,\Phi_{\nu_\mu}(\eta,x)$ is shown.
The histograms are from Monte-Carlo simulations using the SOPHIA code, and
the solid lines correspond to the analytical approximations.

\subsubsection{Electrons and electron antineutrinos}

The electrons and electron antineutrinos are produced through the decay $\mu^-_{}$, which
in its turn is a product of the decay of $\pi^-_{}$-meson. Therefore, for production of
$e^-_{}$ and $\bar{\nu}_e$, at least two pions should
be produced. The production of two pions is energetically allowed if
\be\label{gelec0}
\eta>\eta''_{\min}=4\,r\,(1+r)\approx 2.14 \eta_0^{}\,.
\ee
The maximum and minimum energies of the pion correspondingly are
\be\label{gelec1}
E_{\pi\,\max}=x_{\max}\,E_p\,,\quad E_{\pi\,\min}=x_{\min}\,E_p,,
\ee
where
\be\label{gelec2}
\!\!\! x_{\max\atop\min}=\frac1{2\,(1+\eta)}\left(\eta-2\,r\pm \sqrt{\eta\,(\eta-4\,r\,(1+r))}\right).
\ee
Then $x'_+=x_{\max}$ and $x'_-=x_{\min}/2$.
Eq.(\ref{gelec2}) is obtained from Eq.(\ref{kin7}) if one sets $R=1+r$.
These functions together with
\be
\psi=6\left(1-e^{1.5\,(4-\rho)}\right)\Theta(\rho-4)\,,\quad \rho=\frac{\eta}{\eta_0}\,,
\ee
determine the distributions for $e^-$ and $\bar{\nu}_e$ given in a general form by
Eqs.~(\ref{pggq1})~-~(\ref{pggq1b});
$\Theta(\rho)$ is the Heaviside function ($\Theta(\rho)=0$ if $\rho<0$ and $\Theta(\rho)=1$ if $\rho\ge0$).

\begin{table}
\caption{\label{Table3} Numerical values of parameters $s_l^{}$, $\delta_l^{}$ and $B_l^{}$
for electrons and electron antineutrinos}

\begin{tabular}{c|ccc|ccc}
\hline
$\eta/\eta_0^{}$ &$s_{e^-}$ &$\d_{e^-}$& $B_{e^-}\,,\,
{\rm cm}^3/{\rm s}\phantom{\ds \frac ss}\!$
&$s_{\bar\nu_e}$ &$\d_{\bar\nu_e}$&
$B_{\bar\nu_e}\,,\,{\rm cm}^3/{\rm s}\phantom{\ds \frac ss}\!$\\
\hline
3.0 &0.658&3.09&$6.43\cdot 10^{-19}$&0.985&2.63&$6.61\cdot 10^{-19}\phantom{\ds \frac ss}\!$\\
4.0 &0.348&2.81&$9.91\cdot 10^{-18}$&0.378&2.98&$9.74\cdot 10^{-18}\phantom{\ds \frac ss}\!$\\
5.0 &0.286&2.39&$1.24\cdot 10^{-16}$ &0.31&2.31&$1.34\cdot 10^{-16}\phantom{\ds \frac ss}\!$\\
6.0 &0.256&2.27&$2.67\cdot 10^{-16}$&0.327&2.11&$2.91\cdot 10^{-16}\phantom{\ds \frac ss}\!$\\
7.0 &0.258&2.13&$3.50\cdot 10^{-16}$&0.308&2.03&$3.81\cdot 10^{-16}\phantom{\ds \frac ss}\!$\\
8.0 &0.220&2.20&$4.03\cdot 10^{-16}$&0.292&1.98&$4.48\cdot 10^{-16}\phantom{\ds \frac ss}\!$\\
9.0 &0.217&2.13&$4.48\cdot 10^{-16}$&0.260&2.02&$4.83\cdot 10^{-16}\phantom{\ds \frac ss}\!$\\
10.0 &0.192&2.19&$4.78\cdot 10^{-16}$ &0.233&2.07&$5.13\cdot 10^{-16}\phantom{\ds \frac ss}\!$\\
30.0 &0.125&2.27&$1.64\cdot 10^{-15}$&0.135&2.24&$1.75\cdot 10^{-15}\phantom{\ds \frac ss}\!$\\
100.0 &0.0507&2.63&$4.52\cdot 10^{-15}$&0.0770&2.40&$5.48\cdot 10^{-15}\phantom{\ds \frac ss}\!$\\
\end{tabular}
\end{table}

\section{Photons and leptons produced at interactions of protons with 2.7K CMBR}

In this section we compare the energy spectra of gamma-rays, neutrinos, and
electrons produced at photomeson interactions. For monoenergetic protons interacting with a
radiation field with energy distribution $f_{\rm ph}(\e)$, the energy spectra of photons
and leptons can be reduced to the calculation of a one-dimensional integral
\be
\label{spgl1}
\frac{ dN}{dx}=\int\limits_{\e_0^{}}^{\infty}\! f_{\rm ph}(\e)\;
\Phi\!\left(\eta,\,x\right) d\e\,,
\ee
where $\e_0^{}=\eta_0^{}m_p^2c^4/(4E_p^{})$, and $x=E/E_p$ is the fraction of energy of the
protons transferred to the given type of secondary particle; $\Phi$
is one of the functions described in the previous section.

\begin{figure*}
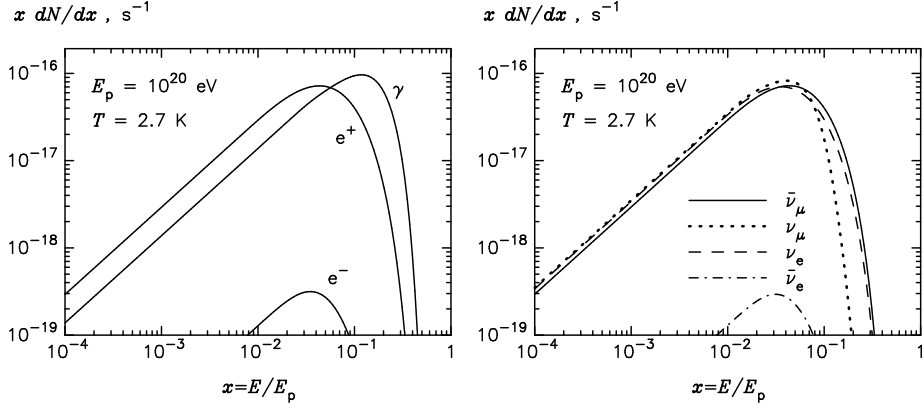

\begin{center}
\mbox{\includegraphics[width=0.3\textwidth,angle=-90]{fig6a.eps}\quad
\includegraphics[width=0.3\textwidth,angle=-90]{fig6b.eps}}
\caption{\label{spec_E20}\small The energy spectra of stable products of
photomeson interactions of a proton of energy $E_{\rm p}=10^{20}\,{\rm eV}$ with the
2.7~K CMBR. Left panel - gamma-rays, electrons, and positrons, right panel - electron
and muon neutrinos and antineutrinos.}
\label{spec_gam20}
\end{center}
\end{figure*}

\begin{figure*}
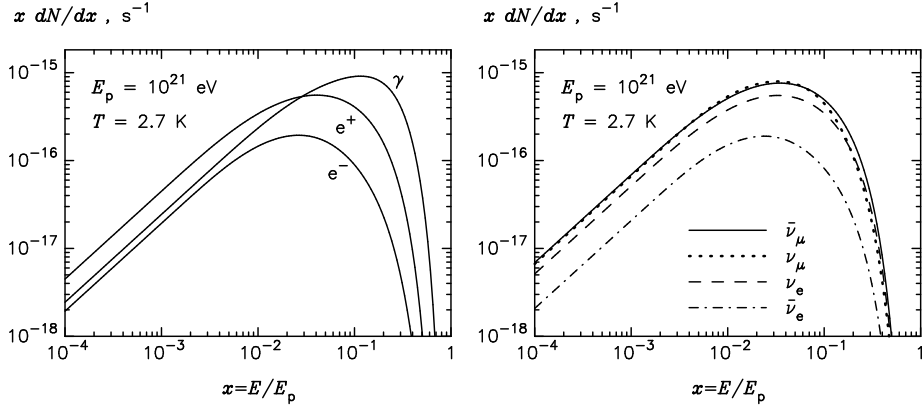

\begin{center}
\mbox{\includegraphics[width=0.3\textwidth,angle=-90]{fig7a.eps}\quad
\includegraphics[width=0.3\textwidth,angle=-90]{fig7b.eps}}
\caption{\label{spec_E21}\small The same as in Fig.~\ref{spec_E20}, but for a proton of
energy $E_{\rm p}=10^{21}\,{\rm eV}$ }
 \label{spec_gam21}
\end{center}
\end{figure*}

In Fig.~\ref{spec_E20} and \ref{spec_E21} we show the energy spectra of
$\gamma$-rays and electrons (left panels) and all neutrino types (right panels)
produced by protons of energy $10^{20}$~eV and $10^{21}$~eV interacting with
blackbody radiation of temperature $T=2.7$~K. The results depend only on the product
$E_p \times T$, therefore they can be easily rescaled to a blackbody radiation of
an arbitrary temperature. The chosen radiation field and proton energies are of
great practical interest in the context of origin and intergalactic propagation of
ultrahigh energy cosmic rays.
Because of interactions with the intergalactic radiation fields, ultrahigh energy
$\gamma$-rays achieve the observer from distances less than 1~Mpc
(see e.g. Ref.~\cite{Coppi}).
The electrons rapidly cool via synchrotron radiation or, in the case of very
small intergalactic magnetic field, initiate electromagnetic cascades
supported by interactions of electrons and gamma-rays with the 2.7~K CMBR.
Only neutrinos freely penetrate through intergalactic radiation and magnetic fields and thus
carry a clear imprint of parent protons.

\begin{figure}
\begin{center}
\includegraphics[width=0.35\textwidth,angle=-90]{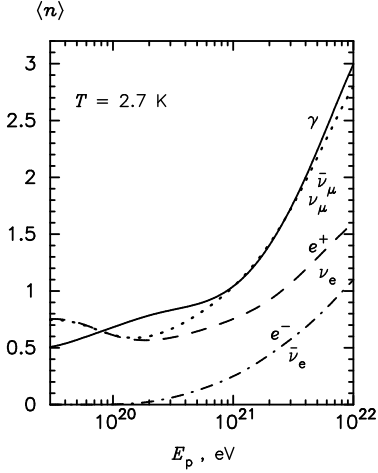}
\caption{\label{multy}\small
The multiplicity of photons and leptons produced in one interaction of a relativistic
proton with 2.7 CMBR}
\end{center}
\end{figure}

In Fig.~\ref{multy} we show the average number of secondaries (multiplicity) produced
in one inelastic interaction of protons with 2.7~K CMBR as a function of proton energy.
The results of numerical calculations are obtained using the
energy spectra of secondary photons, electrons and neutrinos, and the
total cross-section shown in Fig~\ref{gamma_p}. Note that below the threshold of production
of two pions one should have the following relations
\be\label{spgl2}
\frac12\,\langle n_\gamma\rangle+\langle n_{e^+}\rangle=1\,,\quad \langle n_{e^+}\rangle
=\langle n_{\bar\nu_\mu}\rangle=\langle n_{\nu_\mu}\rangle\,.
\ee
The results of calculations based on approximate ana\-ly\-ti\-cal presentations of functions
$\Phi_l$ satisfy these relations with an accuracy of better than 5\%.

\begin{figure}
\begin{center}
\includegraphics[width=0.33\textwidth,angle=-90]{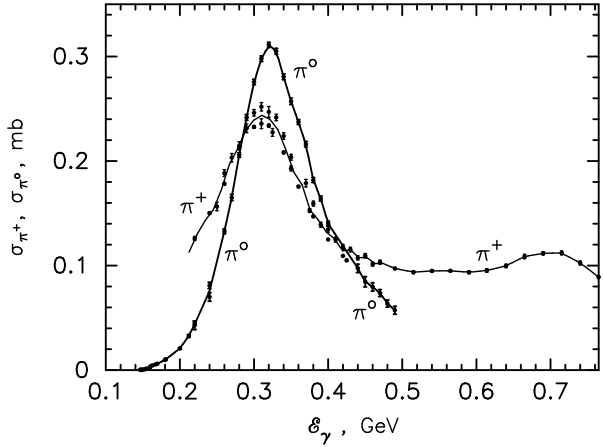}
\caption{\label{g_pi0_plus}\small The total cross-sections of production of
$\pi^+$ and $\pi^0$-mesons as a function of energy of incident gamma-ray
in the rest frame of proton. The experimental points are taken from the site
http:/$\!$/wwwppds.ihep.su:8001.}
\end{center}
\end{figure}

Note that at very low energies the average number of $\gamma$-rays
$\langle n_\gamma\rangle$ appears
smaller than average number of positrons $\langle n_{e^+}\rangle$.
This, at first glance
unexpected result is actually a direct consequence
of the experimental fact that near the threshold the total cross-section
$\sigma_{\pi^+_{}}$ of $\pi^+_{}$ production significantly
exceeds $\sigma_{\pi^0_{}}$ (see Fig.~\ref{g_pi0_plus}).

\section{Production of electron-positron pairs}

At energies below the photomeson production, the main channel
of inelastic interactions for protons with ambient photons
proceeds through the direct production of electron-positron pairs.
In the rest frame of the proton, this process is described by the
so-called Bethe-Heitler cross-section. In astrophysical
environments, the process is more often realized when
ultrarelativistic protons collides with low-energy
photons,
\be\label{pair1}
p+ \gamma \to e^+_{}+e^-_{}+p
\ee
The process is energetically allowed when
\be\label{pair1a}
\gamma_p\e> m_e c^2\,,
\ee
where $\gamma_p=E_p/m_pc^2$ is the proton Lorentz factor,
$\e$ is the soft photon energy, and $m_e$ is the mass of electron.
The maximum energy of the electron (positron) is determined by the kinematics of the process
\be\label{pair1b}
E_{e\max}=\frac{\gamma_p}{1+4\gamma_p\e/(m_pc^2)}\left(\sqrt{\gamma_p\e\mathstrut}+
\sqrt{\gamma_p\e-m_e c^2\mathstrut}\right)^2.
\ee
This equation is valid when
$\gamma_p\gg 1$ and $\e\ll m_p\gamma_p c^2$. In the interval
\be\label{pair1c}
m_e c^2\ll\gamma_p\e\ll m_p c^2
\ee
the maximum electron energy is
\be\label{pair1d}
E_{e\max}=4\,\gamma_p^2\e\,,
\ee
This applies for $E_{e\max}\ll E_p$. In the limit of $\gamma_p\e\gg m_p c^2$
\be\label{pair1e}
E_{e\max}=m_p c^2\gamma_p=E_p\,,
\ee
i.e. the whole energy of the proton is transferred to one of the electrons.

Let's denote by $d\sigma$ the differential cross-section of the process.
The interaction rate is
\be
dw=c^3\,\frac{(k\cdot p)}{\e E_p }\,d\sigma=c^2\,\frac{(k\cdot u_p)}{\e\gamma_p }\,d\sigma\,,
\ee
where $k$ and $p$ are four-momenta of the photon and proton,
$u_p=p/m_pc$ is the four-velocity of the proton,
$(k\cdot p)=\e E_p/c^2-{\bf kp}$ is the scalar product of four-vectors. Let assume
that in a unit volume we have $f_{\rm ph}(\e)\,d\e\,d\Omega/4\pi$ photons between
the energy interval $(\e,\,\e+d\e)$ and moving within the solid angle $d\Omega$.
Then the number of interactions per unit of time is
\be\label{Pair2}
N=c^2\int\! d\e\,\frac{d\Omega}{4\pi}\, f_{\rm ph}(\e)\,\frac{(k\cdot u_p)}{\e \gamma_p}\int d\sigma\,,
\ee
where the integration is performed over all variables.

Below we perform calculations based on the following approach. If we are interested in
a distribution of some variable $\xi$, which is a function $\varphi$ of particle momenta,
this distribution can be found introducing an additional $\d$-function under the integral
in Eq.(\ref{Pair2}):
\be\label{Pair3}
\frac{dN}{d\xi}=c^2\int\! d\e\,\frac{d\Omega}{4\pi}\, f_{\rm ph}(\e)\,\frac{(k\cdot u_p)}{\e \gamma_p}
\int \d(\xi-\varphi)\, d\sigma\, .
\ee
In particular, the energy distribution of electrons in the laboratory frame can be
calculated using the following formula
\be\label{Pair4}
\frac{dN}{dE_e}=c^2\int\! d\e\,\frac{d\Omega}{4\pi}\, f_{\rm ph}(\e)\,\frac{(k\cdot u_p)}{\e\gamma_p}
\int \d(E_e-c(u_{l\!f}^{}\cdot p_e))\, d\sigma\,,
\ee
where $u_{l\!f}^{}$ is the four-velocity of the laboratory frame, and $p_e$ is four-momentum
of electron since the scalar $c(u_{l\!f}^{}\cdot p_e)$ is equal to the energy of electron in the
laboratory system. The proton Lorentz-factor in the laboratory system also can be
considered as a relativistic invariant: $\gamma_p=(u_{l\!f}^{}\cdot u_p)$.

Note that the integral
\be\label{Pair5}
S\equiv\int\,\d(E_e-c(u_{l\!f}^{}\cdot p_e)) \, d\sigma
\ee
is a relativistic invariant, so it can be calculated in any frame of coordinates.
The differential cross-section $d\sigma$ can be written in the simplest form in the
rest-frame of the proton, therefore for calculations of $S$ we will use this system
of coordinate where
\be\label{Pair6}
c(u_{l\!f}^{}\cdot p_e)=\gamma_p(E_--V_p p_-\cos\theta_-)\,.
\ee
Here $E_-^{}$ is the energy and $p_-=\sqrt{E_-^2/c^2-m_e^2c^2}$ is the momentum modulus of electron
in the rest frame of the proton,
$\theta_-$ is the angle between the momenta of photon and electron. Therefore
\be\label{Pair7}
S=\int d\sigma\,\d(E_e-\gamma_p(E_--V_p p_-\cos\theta_-))\,.
\ee
After integration over all variables, except for $E_-$ and $\theta_-$,
the result can be written in the form
\ba\label{Pair8}
&\ds S=\int W(\omega,E_-,\cos\theta_-)\times&\nonumber\\[4pt]
&\d(E_e-\gamma_p(E_--V_p p_-\cos\theta_-))\,dE_-\,d(\cos\theta_-)\,,&
\ea
where
\be\label{Pair9}
W(\omega,E_-,\cos\theta_-)=\frac{d^2\sigma}{dE_-\,d(\cos\theta_-)}
\ee
is the double-differential cross-section as a function of energy and emission angle
of the electron in the rest frame of the proton; $\o=( u_p\cdot k)/(m_ec)$ is the energy of the photon
the rest frame of proton in units $m_ec^2$. The function $W$ has been derived
in the Born approximation in Refs.~\cite{gluck1, gluck2}. The approach used in these
papers describes the production of an electron-positron pair
by a photon in the Coulomb field
which formally corresponds to the limit $m_p\to\infty$. However, we warn the reader
that there is a misprint in the cross-section published in these papers, therefore
we advise to use Eq.(10) of paper by Blumenthal \cite{Blumenthal}, where the typo is fixed.
Note that in Refs.~\cite{Blumenthal,gluck1,gluck2} the system of units is used in which
$c=\hbar=m_e^{}=1$. Since here we cite to certain equations of these papers,
in this section, in order to avoid a confusion, we use the same system of units.

The presence of the $\d$-function in the integrand allows the integration
over the variable $d(\cos\theta_-)$, which gives
\be\label{Pair10}
S=\frac{1}{\gamma_pV_p}\int\!\frac{dE_-}{p_-}\,W(\omega,E_-,\xi)\,,
\ee
where
\be\label{Pair11}
\xi\equiv\cos\theta_-=\frac{\gamma_pE_- -E_e}{\gamma_p V_p p_-}\,.
\ee

After substituting Eq.(\ref{Pair10}) into Eq.(\ref{Pair4}), and using the relation
\be\label{Pair12}
\o=( u_p\cdot k)=\e\gamma_p\,(1-\cos\theta)\,,
\ee
it is convenient to perform the integration over $\o$ instead of integration over the angle.
Then, for ultrarelativistic protons ($\gamma_p\gg1$), we obtain
\ba\label{Pair13}
 &\ds\frac{d{N}}{dE_e}=\frac{1}{2\gamma_p^3}\int
\limits_{\frac{(\gamma_p^{}+E_e)^2}{4\gamma_p^2E_e}}^{\infty}\! d\e\,
\frac{f_{\rm ph}(\e)}{\e^2}\int\limits_{\frac{(\gamma_p^{}+E_e)^2}{2\gamma_p^{}E_e}}
^{2\gamma_p \e}d\omega\,\omega\times\nonumber&\\[4pt]
&\ds \int\limits_{\frac{\gamma_p^2+E_e^2}{2\gamma_p^{}E_e}}^{\omega-1}\frac{dE_-}{p_-}\,W(\omega,E_-,\xi)\, .&
\ea
When substituting Eq.(\ref{Pair11}) into (\ref{Pair13}) we set $V_p=1$,
and correspondingly $\xi=(\gamma_pE_- -E_e)/(\gamma_p p_-)$. The integration
limits in Eq.(\ref{Pair13}) are found from the analysis of kinematics.

In the case of monoenergetic target photon field,
\be
f_{\rm ph}(\e')=C\,\delta(\e'-\e) \,
\ee
the energy distribution of electrons can be written in the form of
the double-integral
\be\label{Pair13a}
 \ds\frac{d{N}}{dE_e}=\frac{C}{2\gamma_p^3\e^2}
\int\limits_{\frac{(\gamma_p^{}+E_e)^2}{2\gamma_p^{}E_e}}
^{2\gamma_p \e}d\omega\,\omega\times\nonumber
\ds \int\limits_{\frac{\gamma_p^2+E_e^2}
{2\gamma_p^{}E_e}}^{\omega-1}\frac{dE_-}{p_-}\,W(\omega,E_-,\xi)\,,
\ee
with the following kinematic condition
\be
4\e\gamma_p^2E_e\ge (\gamma_p^{}+E_e)^2\,.
\ee

For the important case of Planckian distribution of target photons,
\be\label{Pair14}
f_{\rm ph}(\e)=\frac{1}{\pi^2}\,\frac{\e^2}{e^{\e/kT}-1} \ ,
\ee
the expression can be simplified. Indeed, rewriting the first
term in the integrand of Eq.(\ref{Pair13}) in the form
\be\label{Pair15}
d\e\,\frac{f_{\rm ph}(\e)}{\e^2}=\frac{kT}{\pi^2}\,d\ln\!\left(1-e^{-\e/kT}\right),
\ee
we can perform integration over $d\e$ by parts, which after simple transformation leads to the energy spectrum of electrons in the form of double-integral
\ba\label{Pair16}
 &\ds \frac{dN}{dE_e}=-\frac{kT}{2\pi^2\gamma_p^3}
\int\limits_{\frac{(\gamma_p^{}+E_e)^2}{2\gamma_pE_e}}^{\infty}
d\omega\,\omega\ln\!\left(1-e^{-\omega/(2\gamma_p kT)}\right)\times\nonumber&\\[4pt]
&\ds \int\limits_{\frac{\gamma_p^2+E_e^2}{2\gamma_p^{}E_e}}^{\omega-1}\frac{dE_-}{p_-}\,W(\omega,E_-,\xi)\,.&
\ea

In the Born approximation used in \cite{gluck1,gluck2,Blumenthal},
the energy and angular distributions of electrons and positrons are identical, therefore
Eqs.~(\ref{Pair13}) and (\ref{Pair16}) do not distinguish between electrons and positrons.
Let's discus the condition of applicability of Eqs.~(\ref{Pair13}) and (\ref{Pair16}).
In the proton rest system, the cross-section of production of an electron-positron pair
by the proton and in the Coulomb potential coincide for all emission angles of pairs,
when $\o\ll m_p$. In the laboratory frame this is equivalent to the condition
$\e\gamma_p\ll m_p$, which, taking into account Eq.(\ref{pair1d}),
can be written in the form $E_{e\max}\ll E_p$. Thus, the above obtained results
can be applied to production of electrons and positrons when $E_e\ll E_p$.

\begin{figure}
\begin{center}
\includegraphics[width=0.3\textwidth,angle=-90]{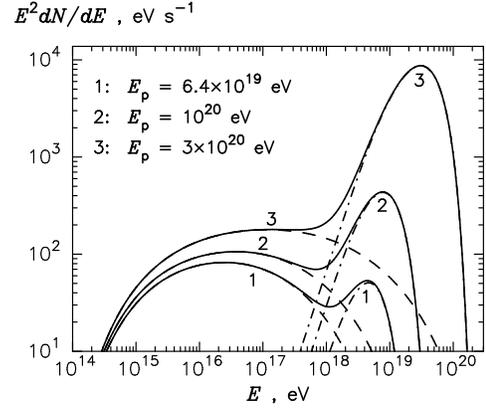}
\caption{\label{pair_photo}\small Energy distributions of electrons and positrons
($N_e=N_+ +N_-$) produced at interactions of protons with 2.7 K CMBR. Dashed lines
and dot-dashed lines
correspond to the pair production and photomeson processes, respectively. Solid lines
show the sums of two contributions. The curves correspond to three energies of protons:
$6.4\times10^{19}\,{\rm eV}$, $10^{20}\,{\rm eV}$ and
$3\times10^{20}\,{\rm eV}$.}
\end{center}
\end{figure}

In Fig.~\ref{pair_photo} we show the energy distributions of electrons and positrons
produced in interactions of protons of three different
energies with the 2.7 K CMBR: $6.4\times10^{19}\,{\rm eV}$, $10^{20}\,{\rm eV}$ and
$3\times10^{20}\,{\rm eV}$. Note that the energy of primary proton
$6.4\times10^{19}\,{\rm eV}$ is interesting in the sense that at this energy
the loss-rates of protons, $E_p^{-1} dE_p/dt$,
due to pair production and photomeson production are equal.
The spectral energy distribution of electron and positrons from the pair production process,
$E^2 dN/dE$, has a bell-type shape with a broad maximum around
$(m_e/m_p) R_p \sim 10^{-3} E_p$. This spectrum is quite different from the
$dN/dE \propto E^{-7/4}$ type energy dependence as it was hypothesized in \cite{Armengaud}.
Fig.~\ref{pair_photo} demonstrates that while the low energy range of electrons
(positrons) is dominated by the
process of pair production, at higher energies the main contribution
comes from photomeson processes. Fortunately, in the energy range where Eqs.~(\ref{Pair13}) and
(\ref{Pair16}) are not valid, the contribution of pair-production to the spectrum of electrons
is negligible compared to the contribution of photomeson processes.

It should be noted that the $\delta$-functional ap\-pro\-xi\-ma\-tion,
which is often used for qualitative estimates of characteristics of
products of high energy interactions, in this specific process
does not provide adequate accuracy.
The reason is that the electrons
produced in a single act of interaction have
very broad energy distribution. The calculations show that
the $\delta$-functional approximation leads to significant
deviation from exact result given by Eq.(\ref{Pair13}), even when one takes into account
the energy dependence of the average fraction of the proton
energy transferred to the electron.

\subsection{Energy losses}

The analytical presentations of the energy spectra of
stable products of interactions of protons with ambient low energy photons
allows us to calculate the energy losses of protons in a radiation field
with arbitrary energy distribution,
\be\label{Loss1}
\frac1{E_p}\,\left|\frac{dE_p}{dt}\right|=\int\limits_0^1\!dx\,x
\int\limits_{\e_{\min}}^\infty\!d\e\, f_{\rm ph}(\e)\,\Phi\!\left(
\eta,\,x\right),
\ee
where $\e_{\min}=\eta_0^{}\,m_p^2c^4/(4E_p)$, and $\Phi(\eta,x)$ is the sum of all seven
energy distribution (relevant to $\gamma$, $e^+$, $e^-$, $\nu_\mu$, $\bar{\nu_\mu}$, $\nu_e$, $\bar{\nu_e}$)
derived in Sec.2. Eq.(\ref{Loss1}) describes the average energy losses transferred to
gamma-rays and leptons. In order to calculate the energy losses due to pair-production
one should multiply Eq.(\ref{Pair16}) to $2E_e$ and integrate over $dE_e$.

\begin{figure}
\begin{center}
\includegraphics[width=0.35\textwidth,angle=-90]{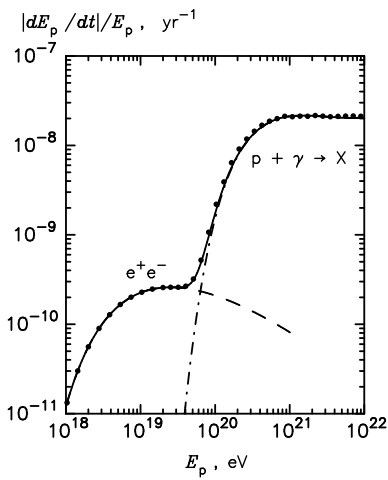}
\caption{\label{ener_loss1}\small The average energy lose rates of protons in the CMBR
with temperature 2.726 K. The lines are obtained using our method of integration of energy
spectra of all final (stable) secondaries, the points are from \cite{Stanev}.
They are obtained from Monte Carlo simulations of interactions of protons with the CMBR
photons using the SOPHIA code. The dashed and dash-dotted lines describe the energy
losses due to pair-production and photomeson interactions, respectively, the solid lines
represent the sum of these two contributions.}
\end{center}
\end{figure}

Calculations of energy losses of protons can be performed
directly, without intermediate calculations of energy distributions of secondary photons
and leptons. In this regard, the energy losses of protons in 2.7 K CMBR have been studied
in great details by many authors, in particular by Berezinsky and coauthors \cite{Berezinsky}
based on a semi-analytical method of calculations and Stanev et al. \cite{Stanev}
based on Monte Carlo simulations using the SOPHIA code. Therefore it is interesting
to compare our results with direct calculations of energy losses. In Fig.~\ref{ener_loss1}
we show the energy lose rate of protons in the blackbody radiation field with temperature
$T=2.726\,{\rm K}$. For comparison, we show the result of calculations
performed using the code SOPHIA \cite{Stanev}.
The agreement of two calculations is an indirect
test of a good accuracy of the obtained above approximate analytical presentations
for energy distributions of stable products from proton-photon interaction.

\begin{figure}
\begin{center}
\includegraphics[width=0.35\textwidth,angle=-90]{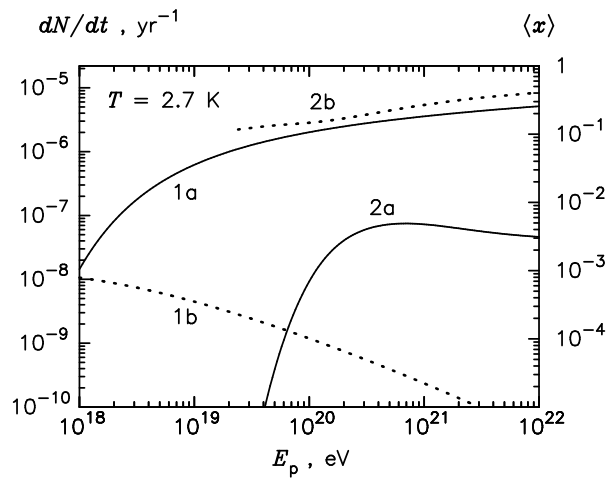}
\caption{\label{events}\small The interaction rates of protons with photons of 2.7~K CMBR
(left axis) and the coefficient of inelasticity (right axis). The curves 1a and 2a are the
electron-positron and photomeson production rates, respectively. The curves 1b and 2b are
the average energy lost by a proton of given energy due to pair production and photomeson
production, respectively.}
\end{center}
\end{figure}

In Fig.~\ref{events} we show the interaction rate of protons with 2.7 K CMBR, as well as the
fraction of energy lost by the proton per interaction (the so-called inelasticity coefficient).
Close to the threshold of pair production around $E \simeq 10^{18}$~eV,
$\langle x\rangle_{e^+e^-}=2m_e/m_p\approx 1.1\cdot 10^{-3}$, as it expected from the
kinematics of the process. However, with an increase of energy, $\langle x\rangle_{\pi}$
gradually decreases down to $10^{-4}$ at $10^{20}$~eV. This effect has been noticed
also in \cite{Chod,Masti}. In the case of photomeson production the
inelasticity coefficient has a quite
different behavior. At the threshold, $\langle x\rangle_{\pi}$
increases from the value of $m_\pi/(m_p+m_\pi)\approx 0.13$ to approximately
0.4 at energy $10^{22}\,{\rm eV}$. Therefore, despite the fact that the cross-section of
pair production significantly, by two orders of magnitude, exceeds the cross-section of
photomeson production, the energy losses at high energies are dominated by
photomeson interactions.

\section{Calculations for power-law distribution of protons}

Instead of integrating Eqs.~(\ref{eq10}) and (\ref{lepton1}) over $d\e$, it is
more convenient to perform integration of these equations over $d\eta$.
This allows the spectra of photons and leptons to be presented in the form
\be\label{eta_dist1}
\frac{dN}{dE}=\int\limits_{\eta_0}^{\infty}H(\eta,E)\,d\eta\,.
\ee
Here
\be\label{eta_dist2}
H(\eta,E)=\frac{m_p^2c^4}{4}\!\int\limits_{E}^{\infty}\!\frac{dE_p}{E_p^2}\,
f_{p}(E_p)\, f_{\rm ph}\!\left(\frac{\eta m_p^2c^4}{4E_p}\right)\!
\Phi\!\left(\eta,\frac{E}{E_p}\right)\!,
\ee
where $E$ is the energy of $\gamma$-rays or leptons, and $\Phi$ is the energy distribution of
the given type of particle.

For photomeson interactions, it is useful to introduce the following
characteristic energy of proton
\be\label{prot_photo3}
E_*=m_pc^2\left(\frac{m_pc^2}{4kT}\,\eta_0^{}\right)\approx 3.0\cdot 10^{20}\;{\rm eV}\,.
\ee
At energy $E_p=E_*$, the proton and a photon of energy $kT$
can produce a pion through a head-on collision.

\begin{figure}
\begin{center}
\includegraphics[width=0.28\textwidth,angle=-90]{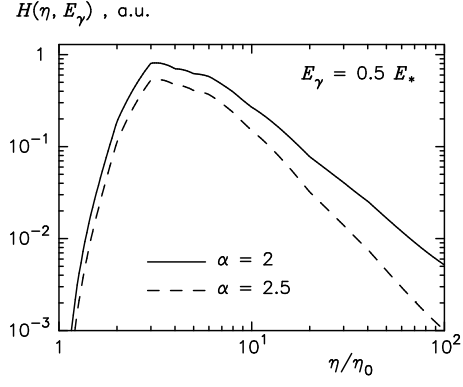}
\caption{\label{eta_dist}\small Function $H(\eta,E_\gamma)$
at fixed energy of $\gamma$-rays, $E_\gamma =1/2 E_\star$, calculated
for a power-law distribution of protons
$f_p(E_p) \propto E_p^{-\alpha}$ with $\alpha=2$ and 2.5. }
\end{center}
\end{figure}

The function $H(\eta,E)$ at fixed energy $E$ describes distribution over $\eta$.
For a power-law distribution of protons, $f_p(E_p)\propto E_p^{-\alpha}$,
the function $H(\eta,E)$ has a maximum at
$\eta/\eta_0\approx 3\,E/E_*$; the position of the maximum slightly depends on the power-law
index $\alpha$ of the proton distribution. The function $H(\eta,E)$ for $\gamma$-rays
is shown in Fig.~\ref{eta_dist} at $E_\gamma=0.5\,E_\star$ and two power-law indices,
$\alpha=2$ and 2.5. At low energies the function $H(\eta,E)$ drops very quickly.
A cutoff in the spectrum of protons results to faster decrease of $H(\eta,E)$ and
to a shift of the point of maximum towards smaller $\eta$.
For electrons and neutrinos $H(\eta,E)$ has similar behavior - a maximum at
$\eta\ll 100\,\eta_0$ and strong decrease with increase of $\eta$.

To use the analytical parameterizations for distributions of photons and electrons 
given by Eq.(27) and Eq.(31) and related Tables I,II,III, which are applicable
for $\eta/\eta_0\le 100$, the following condition should be fulfilled: $ 3\,E/E_*\ll 100$. In this case the main
contribution to the integral (\ref{eta_dist1}) comes from the region
$\eta\ll 100\,\eta_0$, i.e. from events close to threshold. Therefore
the obtained approximate analytical presentations allow calculations of distribution of
particles in the energy range $E\lesssim E_*$.

Finally, let's discuss the production of photons and leptons at interactions of
photons with a realistic distribution of protons, namely a  power-law with an exponential cutoff:
\be\label{prot_photo1}
f_p(E_p)=A E_p^{-2} \exp\left(-\frac{E_p}{E_{\rm cut}}\right)\,.
\ee
where the normalization coefficient is determined from the condition
\be\label{prot_photo2}
\int\limits_{\rm1 GeV}^{\infty}E_p\,J_p(E_p)\,dE_p=1\;\frac{\rm erg}{\,\rm cm^{3}}\,.
\ee

\begin{figure*}
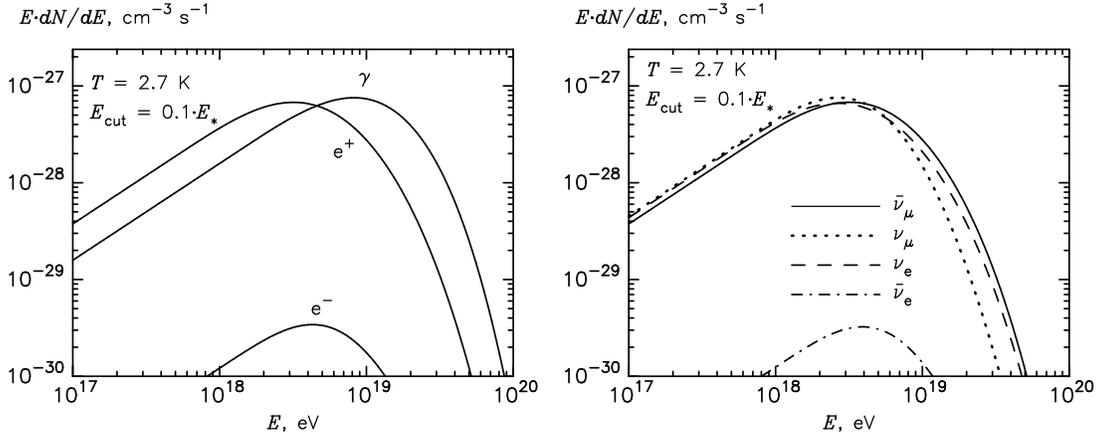

\begin{center}
\mbox{\includegraphics[width=0.32\textwidth,angle=-90]{fig14a.eps}\quad
\includegraphics[width=0.32\textwidth,angle=-90]{fig14b.eps}}
\caption{\label{d_01gam}\small The production spectra ($E dN/dE$) of photons and electrons
(left panel) and neutrinos (right panel) produced with energy distribution described
by Eq.(\ref{prot_photo1}) through the photomeson channel. The cutoff energy in the proton
spectrum is assumed $E_{\rm cut}=0.1 E_*$.}
 \end{center}
\end{figure*}

\begin{figure*}
\begin{center}
\mbox{\includegraphics[width=0.32\textwidth,angle=-90,]{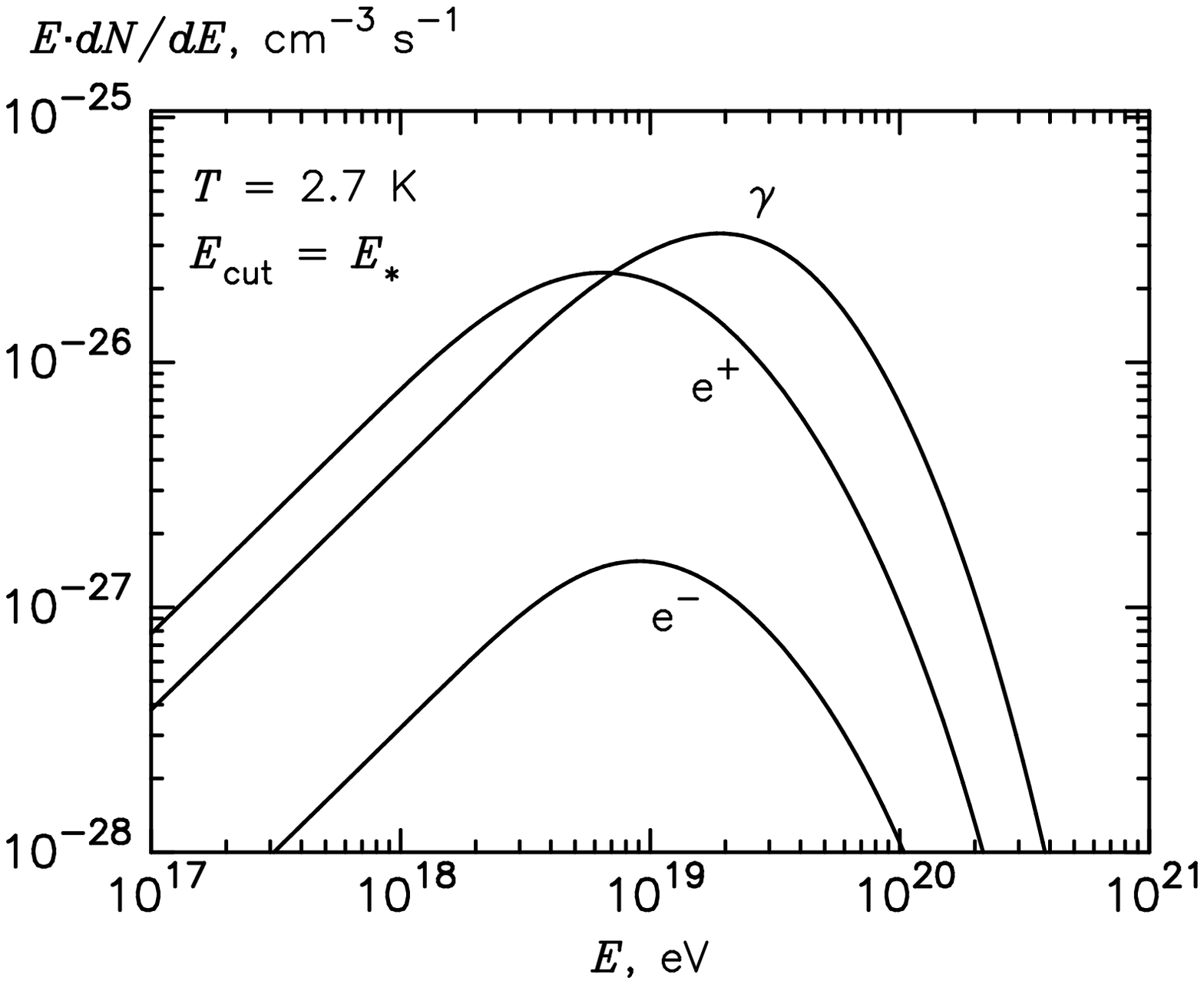}\quad
\includegraphics[width=0.32\textwidth,angle=-90]{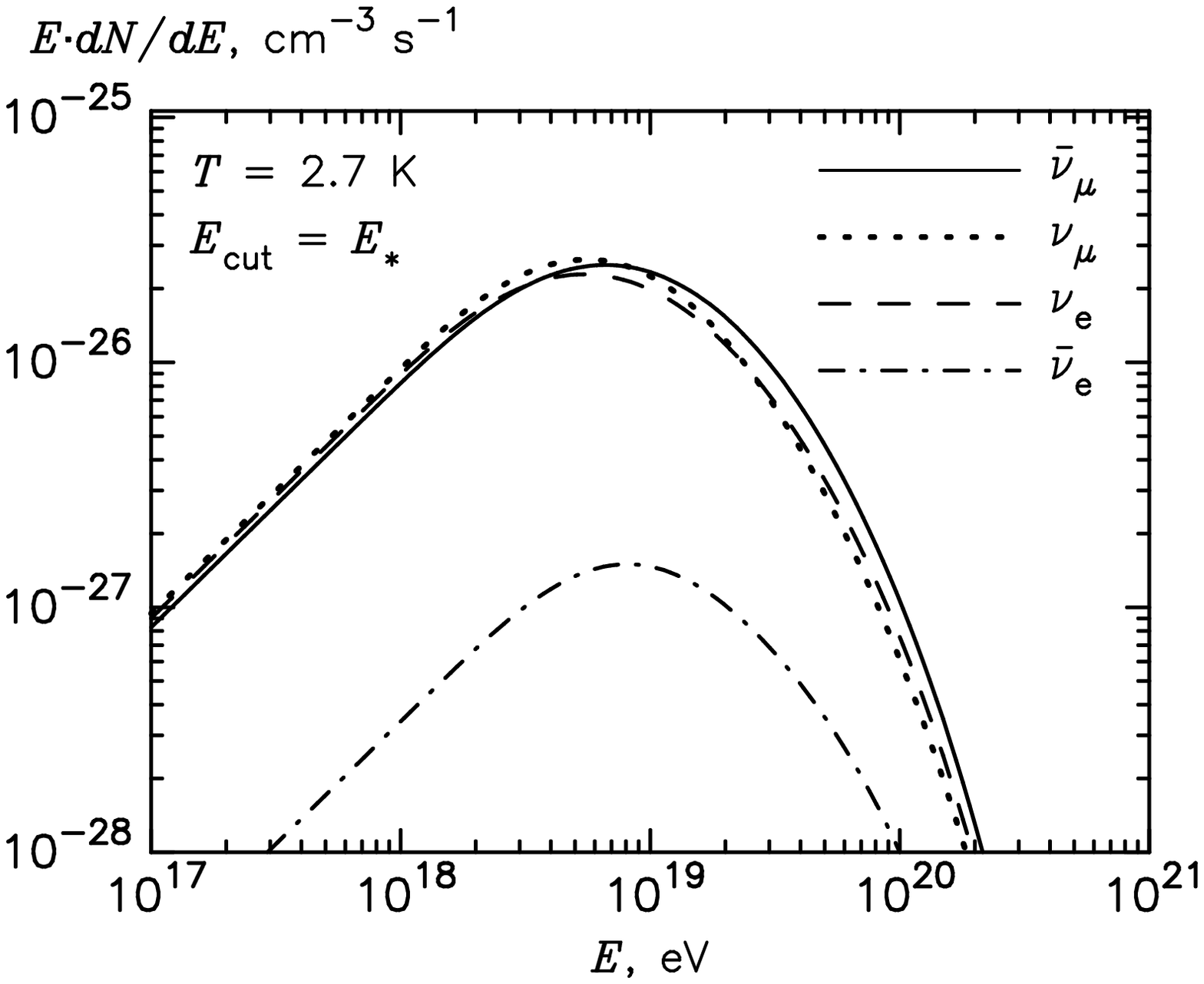}}
\caption{\label{d_1gam}\small The same as in Fig.~\ref{d_01gam}
but for $E_{\rm cut}= E_*$.}
\end{center}
\end{figure*}

\begin{figure*}
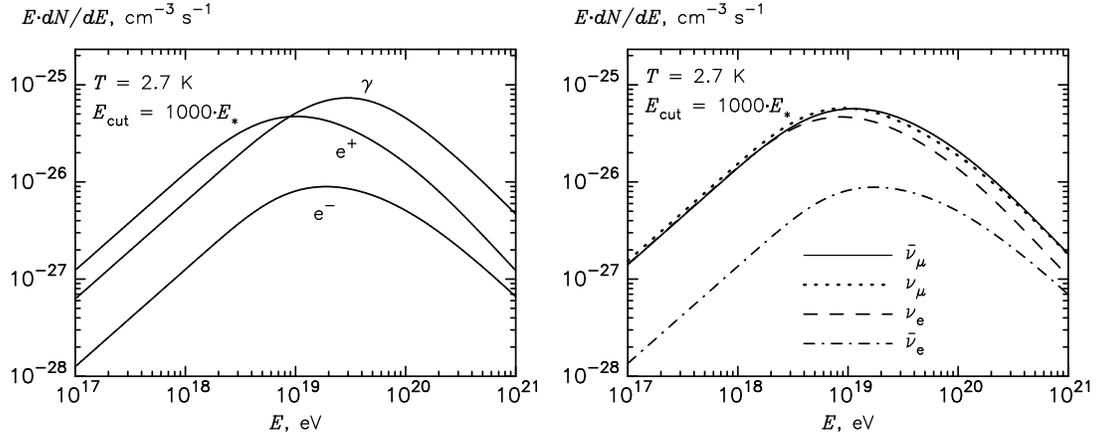

\begin{center}
\mbox{\includegraphics[width=0.32\textwidth,angle=-90]{fig17a.eps}\quad
\includegraphics[width=0.32\textwidth,angle=-90]{fig17b.eps}}
\caption{\label{sd_1000gam}\small The same as in Fig.~\ref{d_01gam}
but for $E_{\rm cut}=1000\cdot E_*$.}
\end{center}
\end{figure*}

In Figs.~\ref{d_01gam} -- \ref{sd_1000gam} we show the spectra of photons, electrons,
and neutrinos produced in photomeson int\-e\-rac\-tions calculated for 4 values of the cutoff
energy in the proton spectrum $E_{\rm cut}=0.1\cdot E_*$, $E_*$, $10\cdot E_*$ and
$10^3\cdot E_*$, respectively. The case $E_{\rm cut}=10^3\cdot E_*$ is almost identical
to a pure power-law spectrum of protons.

\begin{figure*}
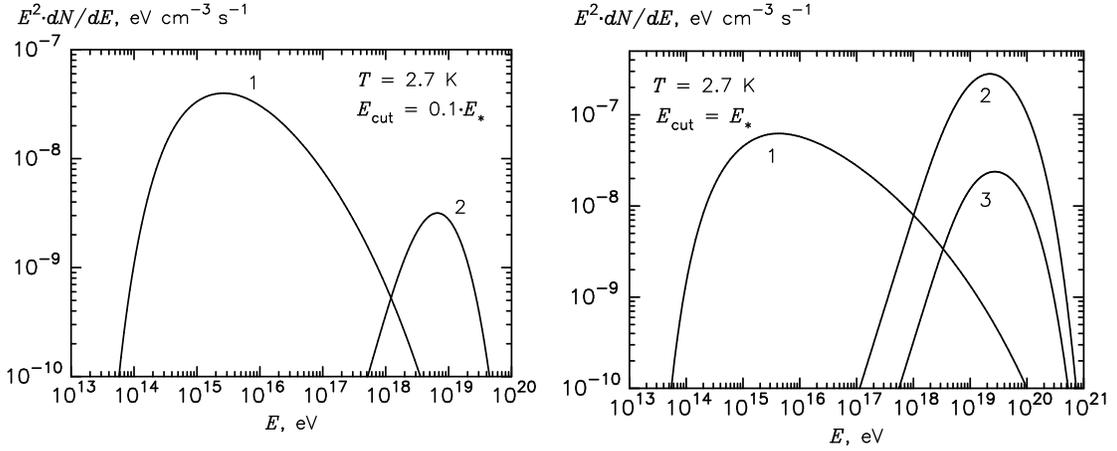

\begin{center}
\mbox{\includegraphics[width=0.32\textwidth,angle=-90]{fig18a.eps}\quad
\includegraphics[width=0.33\textwidth,angle=-90]{fig18b.eps}}
\caption{\label{pair01T}\small The production energy spectra of
electrons and positrons produced
through the channel of pair production (curve 1) and positrons and electrons
produced through the photomeson interactions of protons (curves 2 and 3, respectively).
The proton spectrum is assumed in the form given by Eq.(\ref{prot_photo1})
with cutoff energy at $E_{\rm cut}=0.1 E_*$ (left panel) and $E_{\rm cut}=E_*$
(right panel). Note that the contribution of electrons (curve 3) in left panel 
appears below the low bound of y axis}
\end{center}
\end{figure*}

\begin{figure*}
\begin{center}
\mbox{\includegraphics[width=0.32\textwidth,angle=-90]{fig19a.eps}\quad
\includegraphics[width=0.32\textwidth,angle=-90]{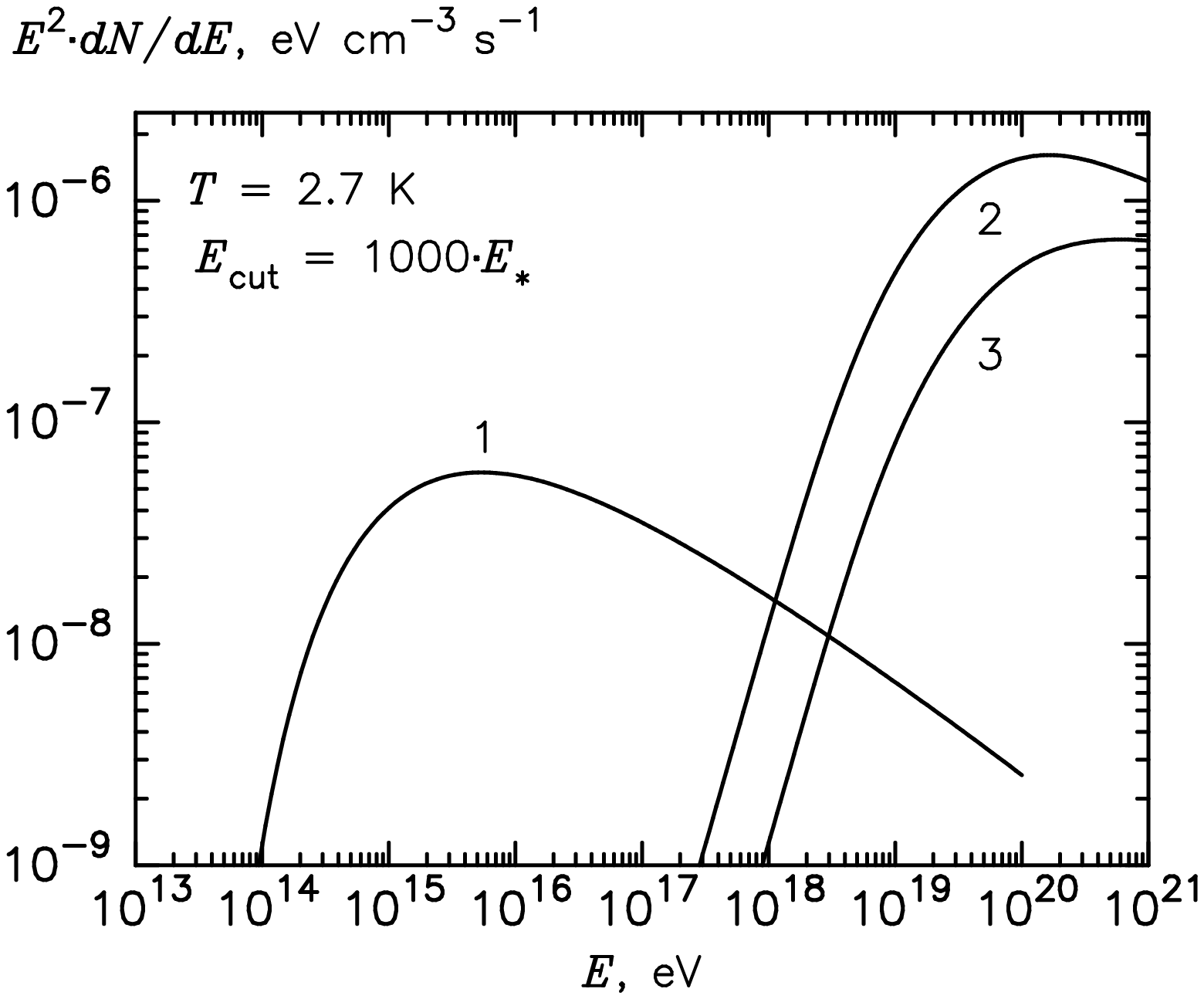}}
\caption{\label{pair10T}\small The same as in Fig.~\ref{pair01T}, but for cutoff energies
$E_{\rm cut}=10 E_*$ (left panel) and $E_{\rm cut}=1000 E_*$
(right panel)}.
\end{center}
\end{figure*}

In Figs.~\ref{pair01T} and \ref{pair10T} we compare the spectra of electrons (and positrons)
produced through the pair production channel with the spectra of electrons from the
decay of photo-produced charged pions.

\begin{figure*}
\begin{center}
\includegraphics[width=0.32\textwidth,angle=-90]{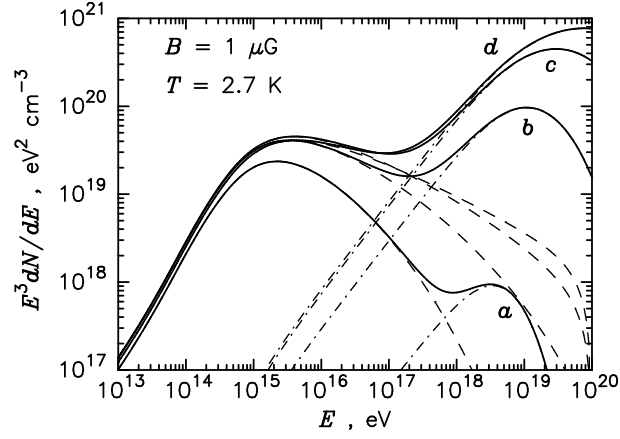}
\caption{\label{cool_e}\small The cooled spectra of electrons and positrons,
$N=N_+ + N_-$. Dashed lines - electrons produced through the pair-production channel,
dot-dashed lines - electrons produced through photomeson interactions. The sum of two
contributions is shown by solid curves.
The proton spectrum is given in the form of Eq.(\ref{prot_photo1}),
with $E_{\rm cut}=0.1\cdot E_*$ $(a)$, $E_*$ $(b)$,
$10\cdot E_*$ $(c)$ and $1000\cdot E_*$ $(d)$.}
\end{center}
\end{figure*}

Finally, in Fig.~\ref{cool_e}, we show the steady-state spectra of
cooled electrons. We assume that electrons are
cooled via synchrotron radiation
in the intergalactic magnetic field $B=1 \ \mu \rm G$
and inverse Compton (IC) scattering on the 2.7 K CMBR.
Since the production spectrum of pair-produced electrons below $E \sim 10^{15}$~eV
drops sharply (see Figs.~\ref{pair01T} and \ref{pair10T}), the synchrotron and
IC cooling (in the Thomson regime) leads
to the formation of a standard $E^{-2}$ type spectrum. This is clearly seen in Fig.~\ref{cool_e}.

\begin{figure*}
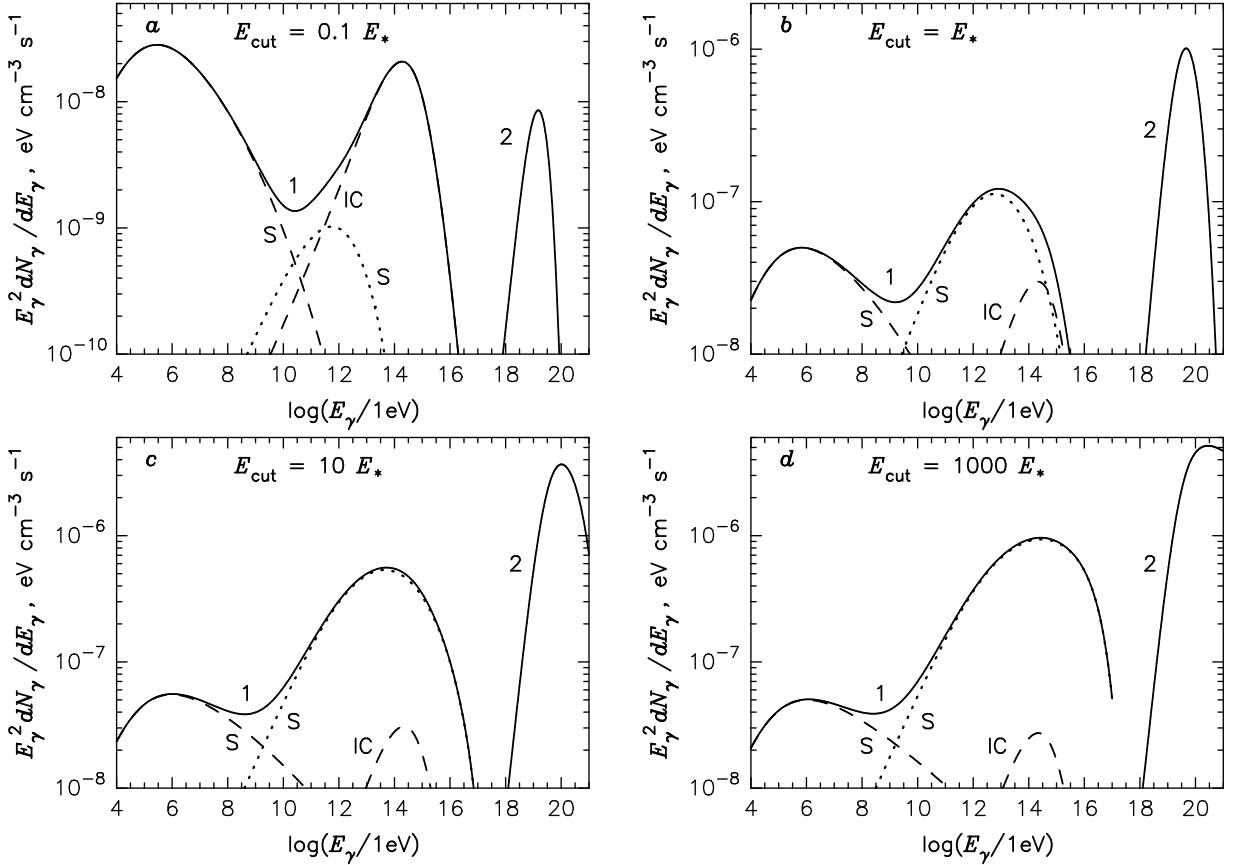

\begin{center}
\mbox{\includegraphics[width=0.32\textwidth,angle=-90]{fig21a.eps}\qquad
\includegraphics[width=0.32\textwidth,angle=-90]{fig21b.eps}}
\mbox{\includegraphics[width=0.32\textwidth,angle=-90]{fig21c.eps}\qquad
\includegraphics[width=0.32\textwidth,angle=-90]{fig21d.eps}}
\caption{\label{syn}\small The synchrotron and IC spectra of cooled electrons.
Dashed line corresponds to synchrotron (S) and IC radiation
of electrons and positrons produced in the pair-production process,
the dotted line corresponds to radiation of positrons (electrons) produced
through photomeson interactions.
Curve 1 (solid line) is the sum of these contributions.
Curve 2 represents the spectrum of gamma-rays
produced at decay of photoproduced $\pi^0$-mesons.
The proton spectrum is given in the form of Eq.(\ref{prot_photo1}),
with $E_{\rm cut}=0.1 E_*$ ($a$), $E_{\rm cut}=E_*$ ($b$),
$E_{\rm cut}=10\,E_*$ ($c$) and $E_{\rm cut}=10\,E_*$ ($d$).
Magnetic field $B=1\,\mu{\rm G}$, temperature $T=2.7\,{\rm K}$.}
\end{center}
\end{figure*}

In Fig.~\ref{syn} we show the spectra of synchrotron and IC radiation
of secondary electrons produced via pair-production and photomeson production channels
for a fixed magnetic field of $B=1 \ \mu \rm G$, the temperature of CMBR $T=2.7$~K
and for 4 different cutoff energies in the proton spectrum $E_{\rm cut}$.
Fig.~\ref{syn}a corresponds to the cutoff energy of protons $E=0.1 E_*$.
In this case the electrons are contributed mainly from pair-production process
with a maximum in the energy distribution ($E^2 dN/dE$) at
energy $E \sim 10^{15} \ \rm eV$. While synchrotron
radiation of these electrons peaks at $E_\gamma \propto B E^ 2\sim 10^6 \ \rm eV$,
the maximum of the IC radiation appears, because of the Klein-Nishina effect,
at $E_\gamma \sim E_\pm \sim 10^{14} - 10^{15} \ \rm eV$. Note that
although energy density of the magnetic field corresponding to $B=1 \mu \rm G$ is
$B^2/8 \pi \approx 4 \times 10^{-14} \ \rm erg/cm^3$, i.e. an order of magnitude
smaller than the energy density of 2.7~K CMBR, emissivity of the synchrotron and IC
components are comparable. This is also a direct consequence of the reduction of the
cross-section of IC scattering of $10^{15} \ \rm eV$ electrons in the Klein-Nishina regime.
The second component of synchrotron radiation related to the
electrons from photomeson processes peaks at much higher energies,
$E_\gamma \sim 10^{12} \ \rm eV$, however its contribution is not significant
because of suppression of the protons at energies above the
threshold of photomeson reactions. The increase of the cutoff energy
in the proton spectrum, $E_{\rm cut}$,
leads to dramatic, orders of magnitude, increase of
the emissivity of the synchrotron radiation of photomeson electrons (see
Figs.~\ref{syn}b,c,d). At the same time, because of the Klein-Nishina cross-section,
only pair-produced electrons contribute to the IC radiation. Therefore
the cutoff energy $E_{\rm cut}$ does not have any
impact on the IC spectrum and emissivity, as it is seen in Figs.~\ref{syn}.

\section{Summary}

We present simple analytical
parametrizations for energy distributions of photons, electrons, and neutrinos
produced in interactions of relativistic protons with
an isotropic monochromatic radiation field. The results
on photomeson processes are obtained using numerical simulations
of proton-photon interactions based on the public available Monte-Carlo
code SOPHIA. We also developed a simple formalism for calculations of
energy spectra of electrons and positron from the pair production
(Bethe-Heitler) process based on the well-known
differential cross-section in the rest frame of the proton.
The energy loss-rate of protons due to photomeson
and pair-production processes in the 2.7~K CMBR calculated by
integrating the energy distributions of the stable products of
interactions is in excellent agreement
with results of previous works based on direct calculations of energy losses
(without intermediate stage of energy distributions of secondaries).
The analytical presentations of energy distributions of photons and leptons
obtained in this paper provide a simple but accurate approach for calculations
of broad-band energy spectra of gamma-rays, electrons and neutrinos in
different astrophysical environments.

\begin{figure*}
\begin{center}
\mbox{\includegraphics[width=0.32\textwidth,angle=-90]{fig16a.eps}\quad
\includegraphics[width=0.32\textwidth,angle=-90]{fig16b.eps}}
\caption{\label{d_10gam}\small The same as in Fig.~\ref{d_01gam}
but for $E_{\rm cut}=10\cdot E_*$.}
\end{center}
\end{figure*}

\section{Acknowledgments}

We appreciate very much the help of Slava Bugayov who provided us with
energy distributions of secondary particles produced at photomeson interactions
based on Monte Carlo simulations using the public available code SOPHIA.
We thank Anita Reimer for providing us with numerical results
of energy losses of protons in the 2.726K CMBR published in \cite{Stanev}.
We thank Andrew Taylor and Mitya Khangulyan for discussion of the results.
Finally we thank the referee for the constructive and very 
useful comments which helped us to improve the paper significantly.

\end{document}